\documentclass[journal]{IEEEtran}
\usepackage{graphics} 
\usepackage{graphicx} 
\usepackage{epsfig} 
\usepackage{mathptmx} 
\usepackage{times} 
\usepackage{amsmath} 
\usepackage{amssymb}  
\usepackage{cite}
\usepackage{lipsum}
\usepackage{adjustbox}
\usepackage{multirow}
\usepackage{hyperref}
\usepackage{textcomp}
\usepackage{algorithmicx}
\usepackage[ruled]{algorithm}
\usepackage{algpseudocode}
\usepackage{xcolor}
\algrenewcommand\alglinenumber[1]{\tiny #1:}

\begin{document}
\bstctlcite{IEEEexample:BSTcontrol}
\title{Learning Based Segmentation of CT Brain Images: Application to Post-Operative Hydrocephalic Scans}

\author{Venkateswararao Cherukuri$^{1,2}$, Peter Ssenyonga$^{4}$, Benjamin C. Warf$^{4,5}$, Abhaya V. Kulkarni$^{6}$, \\
Vishal Monga$^{1, \dag}$, Steven J. Schiff$^{2,3, \dag}$ 
\thanks{*This work is supported by NIH Grant number R01HD085853}
\thanks{$^{\dag}$Contributed equally.}
\thanks{$^{1}$Dept. of Electrical Engineering, $^{2}$Center for Neural Engineering, $^{3}$Dept. Neurosurgery, Engineering Science and Mechanics, and Physics, The Pennsylvania State University, University Park, USA.
        }%
\thanks{$^{4}$CURE Children\textquotesingle s Hospital of Uganda, Mbale, Uganda, $^{5}$Department of Neurosurgery, Boston Children\textquotesingle s Hospital and Department of
Global Health and Social Medicine, Harvard Medical School, Boston, Massachusetts, $^{6}$Division of Neurosurgery, Hospital for Sick Children, University of Toronto, Toronto, ON, Canada
        }
}


\maketitle

\begin{abstract}
\textit{Objective}: Hydrocephalus is a medical condition in which there is an abnormal accumulation of cerebrospinal fluid (CSF) in the brain.  Segmentation of brain imagery into brain tissue and CSF (before and after surgery, i.e.\ pre-op vs. post-op) plays a crucial role in evaluating surgical treatment. Segmentation of pre-op images is often a relatively straightforward problem and has been well researched. However, segmenting post-operative (post-op) computational tomographic (CT)-scans becomes more challenging due to distorted anatomy and subdural hematoma collections pressing on the brain. Most intensity and feature based segmentation methods fail to separate subdurals from brain and CSF as subdural geometry varies greatly across different patients and their intensity varies with time. We combat this problem by a learning approach that treats segmentation as supervised classification at the pixel level, i.e. a training set of CT scans with labeled pixel identities is employed. \textit{Methods}: Our contributions include: 1.) a dictionary learning framework that learns class (segment) specific dictionaries that can efficiently represent test samples from the same class while poorly represent corresponding samples from other classes, 2.) quantification of associated computation and memory footprint, and 3.) a customized training and test procedure for segmenting post-op hydrocephalic CT images. \textit{Results}: Experiments performed on infant CT brain images acquired from the CURE Children\textquotesingle s Hospital of Uganda reveal the success of our method against the state-of-the-art alternatives. We also demonstrate that the proposed algorithm is computationally less burdensome and exhibits a graceful degradation against number of training samples, enhancing its deployment potential.
\end{abstract}


\IEEEpeerreviewmaketitle

\section{Introduction}
\label{sec:Intro}
\subsection{Introduction to the Problem}
\vspace{-.1cm}
Hydrocephalus is a medical condition in which there is an abnormal accumulation of cerebrospinal fluid (CSF) in the brain.
This causes increased intracranial pressure inside the skull and may cause progressive enlargement of the head if it occurs in childhood, potentially causing neurological dysfunction, mental disability and death \cite{adams1965symptomatic}.
The typical surgical solution to this problem is insertion of a ventriculoperitoneal shunt which drains CSF from cerebral ventricles into abdominal cavity. This procedure for pediatric hydrocephalus has failure rates as high as 40 percent in the first 2 years with ongoing failures thereafter \cite{drake1998randomized}. In developed countries, these failures can be treated in a timely manner. However, in developing nations, these failures can often lead to severe complications and even death. To overcome these challenges, a procedure has been developed which avoids shunts known as endoscopic third ventriculostomy and choroid plexus cauterization \cite{warf2007endoscopic}. However, the long-term outcome comparison of these methods has not been fully quantified. One way of achieving quantitative comparison is to compare the volumes of brain and CSF before and after surgery.  These volumes can be estimated by segmenting brain imagery (MR and/or CT) into CSF and brain tissue. Manual segmentation and volume estimation have been carried out but this is tedious and not scalable across a large number of patients. Therefore, automated/semi-automated brain image segmentation methods are desired and have been pursued actively in recent research.

\begin{figure}
 \begin{center}
  \includegraphics[scale=.31]{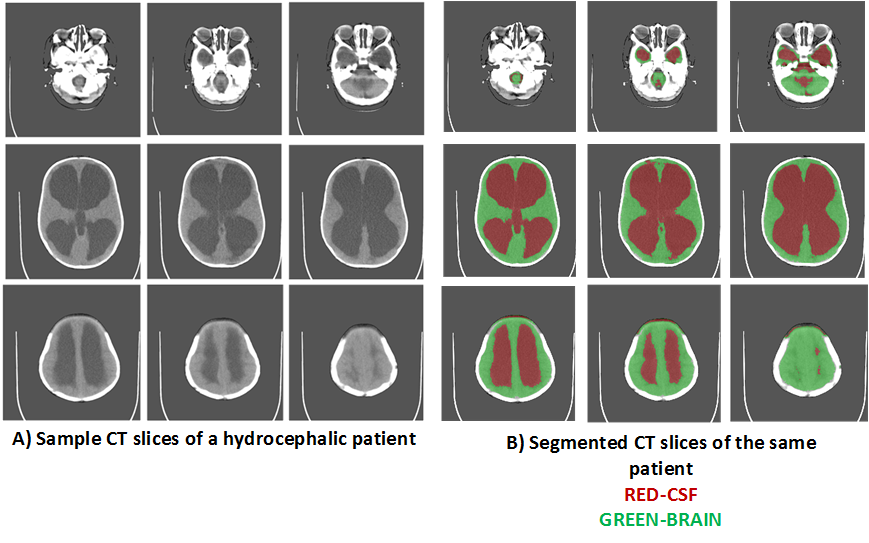}
 \end{center}
 \vspace{-.5cm}
  \caption{A) Sample Pre-operative (pre-op) CT scan slices of a hydrocephalic patient B) Segmented CT-slices of the same patient using \cite{mandell2015volumetric}}
  \label{pre-op}
\end{figure}

\begin{figure}
 \begin{center}
  \includegraphics[scale=.25]{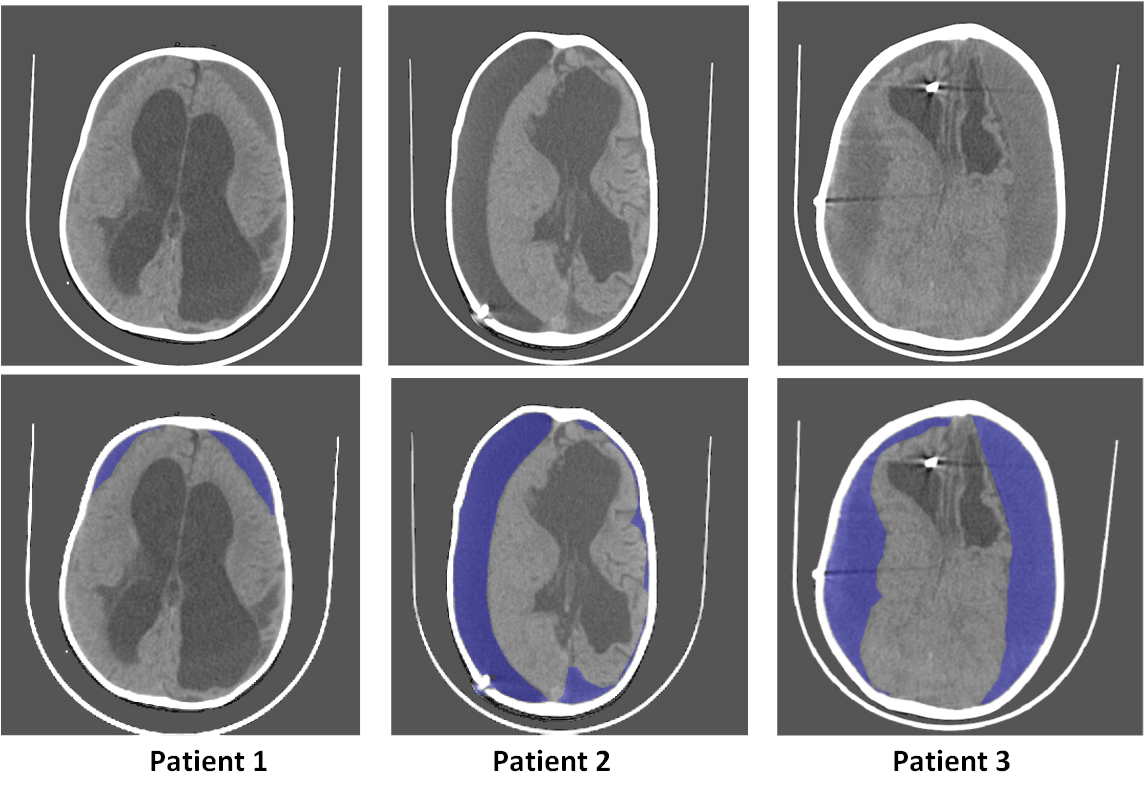}
 \end{center}
  \vspace{-.5cm}
  \caption{Sample post-op CT-images of 3 patients. Top row shows the original images. Bottom row shows subdurals marked in blue. A shunt catheter is visible in patients 2 and 3.}
  \label{subdurals}
\end{figure}

Substantial previous work has been done in the past for segmentation of pre-operative (pre-op) CT-scans of hydrocephalic patients \cite{mandell2015volumetric, mandell2015volumetric2, luo2010wavelet, brandt1994estimation}. It has been noted that the volume of the brain appears to correlate with neurocognitive outcome after treatment of hydrocephalus \cite{mandell2015volumetric2}. Figure \ref{pre-op}A) shows pre-op CT images and Figure \ref{pre-op}B) shows corresponding segmented images using the method from \cite{mandell2015volumetric} for a hydrocephalic patient. The top row of Figure \ref{pre-op}A) shows the slices near base of the skull, second row shows the middle slices and bottom row shows the slices near top of the skull. As we observe from Figure \ref{pre-op}, segmentation of pre-op images can be a relatively simple problem as the intensities of CSF and brain tissue are clearly distinguishable. However, post-op images can be complicated by addition of further geometric distortions and the introduction of subdural hematoma and fluid collections (subdurals) pressing on the brain. These subdural collections have to be separated from brain and CSF before volume calculations are made. Therefore, the images have to be segmented into 3 classes (brain, CSF and subdurals) and subdurals must be removed from the volume determination. Figure \ref{subdurals} shows sample post-operative (post-op) images of 3 patients having subdurals. Note that the subdurals in patient-1 are very small compared to the subdurals in other two patients. Further, large subdurals are observed in patient-3  on both sides of the brain as opposed to patient-2. The other observation we can make is that the intensity of subdurals in patient-2 is close to the intensity of CSF, whereas the intensity of subdurals in other two patients is close to intensity of brain tissue. The histogram of the pixel intensity of the images remains bi-modal making it further challenging to separate subdurals from brain and CSF.


\vspace{-.5cm}
\subsection{Closely Related Recent Work}
\begin{figure}
 \begin{center}
  \includegraphics[scale=.15]{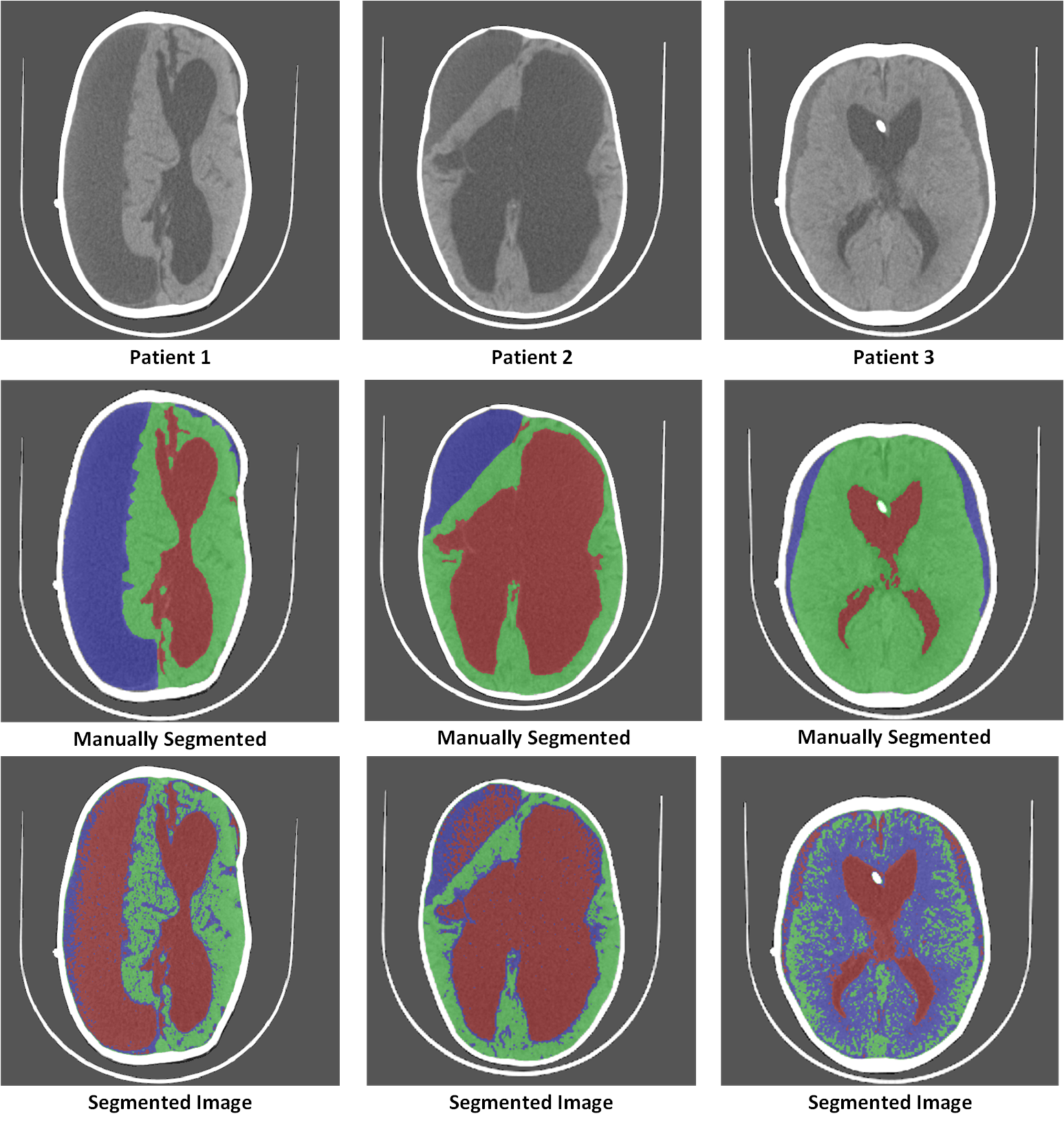}
 \end{center}
  \vspace{-.5cm}
  \caption{Demonstration of segmentation using a traditional intensity based method \cite{bis2014}. Top row represents original images of 3 patients. Second row represents manually segmented images. Third row represents the segmentation using \cite{bis2014}. Green-Brain, Red-CSF, Blue-Subdurals}
  \label{TraditionalDemo}
\end{figure}

\vspace{-.1cm}
Many methods have been proposed in the past for segmentation of brain images \cite{mandell2015volumetric, mayer2009adaptive, li1993knowledge, weisenfeld2009automatic, bis2014, ribbens2014unsupervised, greenspan2006constrained}. Most of these methods work on the principles of intensity based thresholding and model-based clustering techniques. However these traditional methods for segmentation fail to identify subdurals effectively as they are hard to characterize by a specific model, and subdurals pose different range of intensities for different patients. For example, Figure \ref{TraditionalDemo} illustrates the performance of \cite{bis2014} on the images of 3 different patients with subdurals. We can observe that the accuracy in segmenting these images is very poor. Apart from these general methods for brain image segmentation, relatively limited work has been done to identify subdurals \cite{liu2008automatic, liao2009multiresolution, sharma2014classification, gong2013finding, soltaninejad2014hybrid}. These methods work on the assumption that the images have to be segmented into only 2 classes which are brain and subdurals. Therefore, these methods are unlikely to succeed for images acquired from hydrocephalic patients where CSF volume is significant. Because intensity or other features that can help characterize a pixel into one of three segments (brain, CSF and subdurals) are not apparent; they must be discovered via a learning framework.

Recently, sparsity constrained learning methods have been developed for image classification \cite{wright2009robust} and found to be widely successful in medical imaging problems \cite{srinivas2014simultaneous, vu2016histopathological, mousavi2015automated, yu2011group}. The essence of the aforementioned sparse representation based classification (SRC) is to write a test image (or patch) as a linear combination of training images collected in a matrix (dictionary), such that the coefficient vector is determined under a sparsity constraint. SRC has seen significant recent application to image segmentation \cite{wang2014segmentation,wang2014integration,wu2014prostate,zhou2015nuclei, liao2013sparse} wherein a pixel level classification problem is essentially solved.

In the works just described, the dictionary matrix simply includes training image patches from each class (segment). Because each pixel must be classified, in segmentation problems training dictionaries can often grow to be prohibitively large. Learning compact dictionaries  \cite{yang2011fisher,jiang2013label,monga2017handbook} continues to be an important problem. In particular, the Label Consistent K-SVD (LC-KSVD) \cite{jiang2013label} dictionary learning method, which has demonstrated success in image classification has been re-purposed and successfully applied to medical image segmentation \cite{tong2013segmentation, lee2015brain, roy2015subject, bevilacqua2016dictionary, nouranian2016learning}.

\noindent \textbf{Motivation and Contributions:} In most existing work on sparsity based segmentation, a dictionary is used for each voxel/pixel that creates large computational as well as memory footprint. Further, the objective function for learning dictionaries described in the above literature (based invariably on LC-KSVD) is focused on extracting features that characterize each class (segment) well. We contend that the dictionary corresponding to a given class (segment) must additionally be designed to poorly represent out-of-class samples. We develop a new objective function that incorporates an out-of-class penalty term for learning dictionaries that accomplish this task. This leads to a new but harder optimization problem, for which we develop a tractable solution. We also propose the use of a new feature that incorporates the distance of a candidate pixel from the edge of the brain computed via a distance transform. This is based on the observation that subdurals are almost always attached to the boundary of the brain. Both intensity patches as well as the distance features are used in the learning framework. The main contributions of this paper are summarized as follows:
\begin{enumerate}
\item \textbf{A new objective function to learn dictionaries for segmentation under a sparsity constraint:} Because discriminating features are automatically discovered, we call our method feature learning for image segmentation (FLIS). A tractable algorithmic solution is developed for the dictionary learning problem.

\item \textbf{A new feature that captures pixel distance from the boundary of brain} is used to identify subdurals effectively as subdurals are mostly attached to the boundary of the brain. This feature also enables the dictionary learning framework to use a single dictionary for all the pixels in an image as opposed to the existing methods that use a separate dictionary for each pixel type. Incorporating this additional ``distance based feature" helps significantly reduce the computation and memory footprint of FLIS.
\item \textbf{Experimental validation:} Validation on challenging real data acquired from CURE Children\textquotesingle s Hospital of Uganda is performed. FLIS results are compared against manually labeled segmentation as provided by an expert neurosurgen. Comparisons are also made against recent and state of the art sparsity based methods for medical image segmentation.

\item \textbf{Complexity analysis and memory requirements:} We analytically quantify the computational complexity and memory requirements of our method against competing methods. The experimental run time on typical implementation platforms is also reported.
\item \textbf{Reproducibility:} The experimental results presented in the paper are fully reproducible and the code for segmentation and learning FLIS dictionaries is made publicly available at: \url{https://scholarsphere.psu.edu/concern/generic_works/bvq27zn031}.

\end{enumerate}
A preliminary version of this work was presented as a short conference paper at the 2017 IEEE Int. Conference on Neural Engineering \cite{cherukuri2017learning}. Extensions to the conference paper include a detailed analytical solution to the objective function in Eq. (\ref{eq7}). Further, extensive experiments are performed by changing various parameters of our algorithm and new statistical insights are provided. Additionally, a detailed complexity analysis is performed and memory requirements of FLIS along with competing methods is presented.\\
The remainder of the paper is organized as follows. A review of sparsity based segmentation and detailed description of the proposed FLIS is provided in Section \ref{sec:method}. Experimental results are reported in Section \ref{sec:impRes} including comparisons against state of the art. The appendix contains an analysis of the computation and memory requirements of our method and selected competing methods. Concluding remarks are provided in Section \ref{sec:Conc}.

%


\section{Feature Learning For Image Segmentation (FLIS)}
\label{sec:method}

\subsection{Review of Sparse Representation Based Segmentation}
\vspace{-.1cm}
To segment a given image into $C$ classes/segments, every pixel $z$ in the image has to be classified into one of these classes/segments. The general idea is to collect intensity values from a patch of size $w\times w$ (in case of 3D images a patch of size $w\times w\times w$ is considered) around each pixel and to represent this patch as a sparse linear combination of training patches that are already manually labeled. This idea is mathematically represented by Eq. (\ref{eq1}). $m(z)\in \mathbb{R}^{(w^2)\times 1}$ represents a vector of intensity values for a square patch around pixel $z$. $Y(z)\in \mathbb{R}^{(w^2)\times N}$ represents the collection of $N$ training patches for pixel $z$ in a matrix form. $\alpha \in \mathbb{R}^{N\times 1}$ is the vector obtained by solving Eq. (\ref{eq1}). $\parallel\bullet\|_{0}$ represents $l_{0}$ pseudo-norm of a vector which is the number of non-zero elements in a vector. $\parallel\bullet\|_{2}$ represents the $l_{2}$ Euclidean norm.  The intuition behind this idea is to minimize the reconstruction error between $m(z)$ and the linear combination $Y(z)\alpha$ with the number of non-zero elements in $\alpha$ less than $L$. The constraint on $l_{0}$ pseudo-norm hence enforces sparsity. Often the $l_{0}$ pseudo-norm is relaxed to an $l_{1}$ norm \cite{wang2014integration} to obtain fast and unique global solutions.
Once the sparse code $\alpha$ is obtained, pixel likelihood probabilities for each class (segment) $j\in\{1,\ldots,C\}$ are obtained using Eq. (\ref{eq2}) and Eq. (\ref{eq3}). The probability likelihood maps are normalized to 1 and a candidate pixel $z$ is assigned to the most likely class (segment) as determined by its sparse code.

\begin{equation}\label{eq1}
\arg\min_{\|\alpha\|_{0} < L}\parallel m(z) - Y(z)\alpha\parallel_{2}^{2}
\end{equation}

\begin{equation}\label{eq2}
P_{j}(z) = \frac{\sum_{i = 1}^{N}\alpha_{i}\delta_{j}(V_{i})}{\sum_{i = 1}^{N}\alpha_{i}}
\end{equation}
where $V_{i}$ is the $i^{th}$ column vector in the pre-defined dictionary $Y(z)$, and $\delta_{j}(V_{i})$ is an indicator defined as
\begin{equation}\label{eq3}
\delta_{j}(V_{i}) = \begin{cases}1, V_{i} \in \mbox{class }j\\ 0, \mbox{otherwise} \end{cases}
\end{equation}

Note that training dictionaries $Y(z)$ could grow to be prohibitively large, which motivates the design of compact dictionaries that can lead to high accuracy segmentation.  Tong \cite{tong2013segmentation} {\em et al.} adapted the well-known LC-KSVD method \cite{jiang2013label} for segmentation by minimizing reconstruction error along with enforcing a label-consistency criteria. The idea is formally quantified in Eq. (\ref{eq4}). For a given pixel $z$, $Y(z) \in \mathbb{R}^{(w^2)\times N}$ represents all the training patches for pixel $z$. $N$ is the number of training patches. $D(z) \in \mathbb{R}^{(w^2)\times K}$ is the compact dictionary that is obtained with $K$ being the size of the compact dictionary. $\|X\|_{0} < L$, a sparsity constraint means that each column of $X$ has no more than $L$ non-zero elements. $H(z) \in \mathbb{R}^{C\times N}$ represents the label matrix for the training patches with $C$ being the number of classes/segments to which a given pixel can be classified. For example in our case $C = 3$ (Brain, CSF and Subdurals) and the label matrix for a patch around a pixel which has its ground truth as CSF will be $[0 \mbox{ } 1 \mbox{ }0]^{T}$. $W(z) \in \mathbb{R}^{C\times K}$ is the linear classifier which is obtained along with $D(z)$ to represent $H(z)$. $\parallel\bullet\|_{F}$ represents the Frobenius (squared error) norm. The terms in black minimize reconstruction error while the term in red represents the label-consistency criteria. When a new test image is analyzed for segmentation, for each pixel $z$, $D(z)$ and $W(z)$ are invoked and the sparse code $\alpha \in \mathbb{R}^{K\times 1}$ is obtained by solving Eq. (\ref{eq5}) which is an $l_{1}$ relaxation form of Eq. (\ref{eq1}). Unlike the classification strategy used in Eq. (\ref{eq2}), we use the linear classifier $W(z)$ on sparse code $\alpha$ to classify/segment the pixel which is shown in Eq. (\ref{eq6}). Note that $\beta$ is a positive regularization parameter that controls the relative regularization between reconstruction error and label consistency.
\begin{equation}\begin{aligned}\label{eq4}
   \arg \min_{D(z),W(z),X}&\{ \min_{\|X\|_{0} < L}\{\|Y(z)-D(z)X\|_{F}^{2} + \color{red}\beta\|H(z)-W(z)X\|_{F}^{2}\color{black}\}\}
\end{aligned}\end{equation}

\begin{equation}\label{eq5}
\arg\min_{\alpha > 0}\|m(z) - D(z)\alpha\|_{2}^{2} + \lambda\|\alpha\|_{1}
\end{equation}
\vspace{-.3cm}
\begin{equation}\label{eq6}
H_{z} = W(z)\alpha,\mbox{ }label(z) = \arg\max_{j}(H_{z}(j)),\\
\end{equation}

where $H_{z}$ is the class label vector for the tested pixel $z$, and the $\arg\max$ reveals the best labelling achieved through applying $\alpha$ to the linear classifier $W(z)$.

Tong \cite{tong2013segmentation} {\em et al.}'s work is promising for segmentation but we identify two key open problems: 1.) learned dictionaries for each pixel lead to a high computational and memory footprint, and 2.) the label consistency criterion enhances segmentation by encouraging intra- or within-class similarity but inter-class differences must be maximized as well. Our proposed FLIS addresses both these issues.
\vspace{-.25cm}
\subsection{FLIS Framework}
\vspace{-.1cm}
We introduce a new feature that captures the pixel distance from the boundary of the brain. This serves two purposes. First, as we observe from Figure \ref{subdurals}, subdurals are mostly attached to the boundary of the brain. Adding this feature along with the vectorized patch intensity intuitively helps enhance the recognition of subdurals. Secondly, we no longer need to design pixel specific dictionaries because the aforementioned ``distance vector" (for a patch centered around a pixel) provides enough discriminatory nuance.

\noindent \textbf{Notation:} For a given patient, we have a stack of $T$ CT slice images starting from base of the skull to top of the skull which can be observed from Figure \ref{pre-op}. The goal is to segment each image of the stack into three categories: brain, CSF and subdurals.
Let $Y_{B}\in \mathbb{R}^{d\times N_{B}}$, $Y_{F}\in \mathbb{R}^{d\times N_{F}}$ and $Y_{S}\in \mathbb{R}^{d\times N_{S}}$ represent the training samples of brain, CSF and subdurals respectively. Each column of $Y_{i}$, $i\in B,\mbox{ } F,\mbox{ }  S$ represents intensity of the elements in a patch of size $w\times w$ around a training pixel concatenated with the distances from boundary of brain for each pixel in the patch (described in detail in Section \ref{sec:implementation}). $N_{i}$ represents the number of training patches for each class/segment. They are chosen to be same for all the 3 classes/segments.
We denote the dictionaries learned as $D_{i} \in \mathbb{R}^{d\times K}$. $K$ is the size of each dictionary. $X_{i} \in \mathbb{R}^{K\times N_{i}}$ represents the matrix that contains the sparse code for each training sample in $Y_{i}$. $H_{i} \in \mathbb{R}^{3\times N_{i}}$ represents the label matrices of the corresponding training elements $Y_{i}$. For example, a column vector of $H_{B}$ looks like $[1 \mbox{ } 0 \mbox{ }0]^{T}$ and finally, $W_{i}$ denotes the linear classifier that is learned to represent $H_{i}$.
\vspace{-.25cm}
\subsection{Problem Formulation}
\label{sec:formulation}
\vspace{-.1cm}
The dictionary $D_{i}$ should be designed such that it represents in-class samples effectively and poorly represent complementary samples along with achieving the label consistency criteria. To ensure this, we propose the following problem:
    \begin{equation}\begin{aligned}\label{eq7}
   \arg \min_{D_{i},W_{i}}&\bigg\{ \frac{1}{N_{i}}\min_{\|X_{i}\|_{0} < L}\{\|Y_{i}-D_{i}X_{i}\|_{F}^{2} + \color{red}\beta\|H_{i}-W_{i}X_{i}\|_{F}^{2}\color{black}\}\\&- \frac{\rho}{\hat{N_{i}}}\min_{\|\hat{X_{i}}\|_{0} < L}\color{black}\{\color{blue}\|\hat{Y_{i}}-D_{i}\hat{X_{i}}\|_{F}^{2} + \color{brown}\beta\|\tilde{H_{i}}-W_{i}\hat{X_{i}}\|_{F}^{2}\color{black}\}\bigg\}
    \end{aligned}\end{equation}
The terms with $\hat{(\bullet)}$ represent the complementary samples of a given class, $\parallel\bullet\|_{F}$ represents Frobenius norm and  $\parallel X\|_{0} < L$ implies that each column of $\|X\|$ has non-zero elements not more than $L$. The label matrices are concatenated, $\tilde{H_{i}} = [H_{i}\mbox{ } H_{i}]$, to maintain consistency with the dimension of $W_{i}\hat{X_{i}}$, because there are two complimentary samples. $\beta$ and $\rho$ are positive regularization parameters. $\rho$ is an important parameter to obtain a solution for the objective function that we discuss in subsequent sections.

\begin{figure*}
 \begin{center}
  \includegraphics[scale=1]{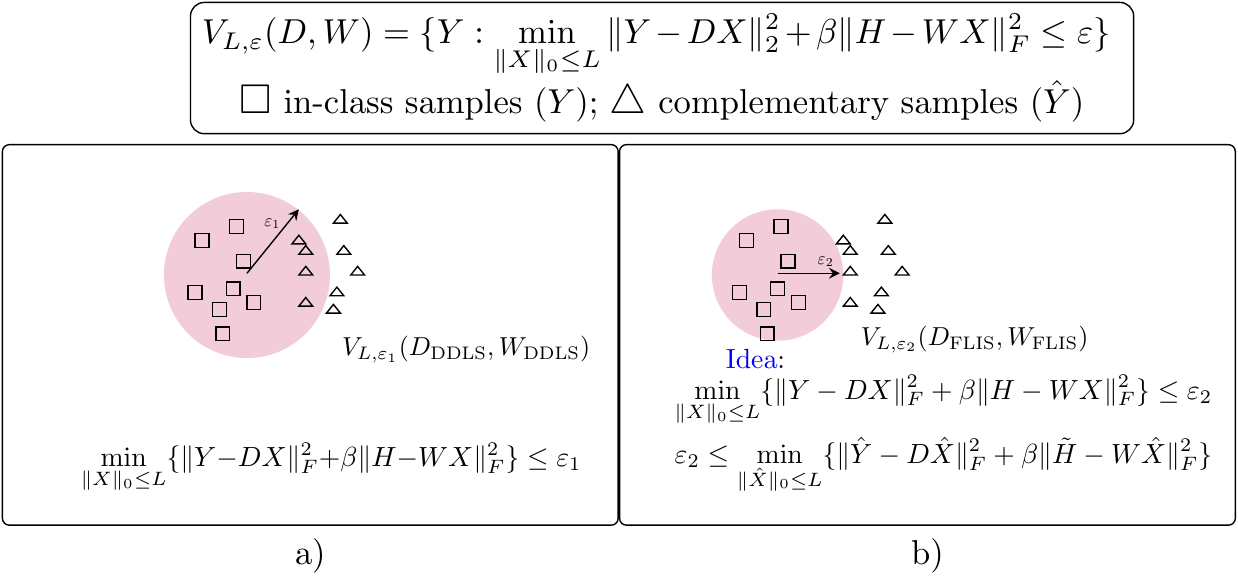}
 \end{center}
  \vspace{-.5cm}
  \caption{Visual representation of our FLIS in comparison with DDLS \cite{tong2013segmentation}. a) represents the idea of DDLS and b) represents a desirable outcome of our idea which is more capable of differentiating in-class and out of class samples.}
  \label{CostFunction}
\end{figure*}

\noindent \textbf{Intuition behind the objective function:} The term in black makes sure that intra-class difference is small and the term in red enforces label-consistency. These two terms make sure that in-class samples are well represented. To represent the complementary samples poorly, the reconstruction error between the complementary samples and the sparse linear combination of in-class dictionary samples should be large. This is achieved through the term in blue. Further, a "label-inconsistency term" is added (in brown) utilizing the sparse code for out of class samples, which again encourages inter-class differences. Essentially, the combination of terms in blue and brown enables us to discover discriminative features that differentiate one class (segment) from another effectively. Note that the objective functions described in \cite{tong2013segmentation,roy2015subject, bevilacqua2016dictionary, nouranian2016learning,lee2015brain} are special cases of Eq. (\ref{eq7}) since they {\em do not} include terms that emphasizes inter-class differences. The visual representation of our idea in comparison with the objective function defined in \cite{tong2013segmentation} (known as discriminative dictionary learning and sparse coding (DDLS)) is shown in Figure \ref{CostFunction}. The problem in Eq. (\ref{eq7}) is non-convex with respect to its optimization variables; we develop a new tractable solution which is reported next.
\vspace{-.25cm}
\subsection{Proposed Solution :}
\label{sec:Solution}
\vspace{-.1cm}
For simplifying notation in Eq. (\ref{eq7}), we replace $Y_{i}$, $\hat{Y_{i}}$, $X_{i}$, $\hat{X_{i}}$, $H_{i}$, $\tilde{H_{i}}$, $W_{i}$, $\hat{W_{i}}$, $N_{i}$, $\hat{N_{i}}$ with $Y$, $\hat{Y}$, $X$, $\hat{X}$, $\hat{X}$, $H$, $\tilde{H}$, $W$, $\hat{W}$, $N$, $\hat{N}$ respectively. Therefore, the cost function becomes
    \begin{equation}\begin{aligned}\label{eq8}
   \arg \min_{D,W}\bigg\{ &\frac{1}{N}\min_{\|X\|_{0} < L}\{\|Y-DX\|_{F}^{2} + \beta\|H-WX\|_{F}^{2}\}\\&- \frac{\rho}{\hat{N}}\min_{\|\hat{X}\|_{0} < L}\{\|\hat{Y}-D\hat{X}\|_{F}^{2} + \beta\|\tilde{H}-W\hat{X}\|_{F}^{2}\}\bigg\}
    \end{aligned}\end{equation}

First, an appropriate $L$ should be determined. We begin by learning an ``initialization dictionary" using the well-known online dictionary learning (ODL) \cite{mairal2010online} given by:
\begin{equation}\label{eq9}
(D^{(0)}, X^{(0)}) = \arg\min_{D,X}\{\|Y-DX\|_{F}^{2} + \lambda\|X\|_{1}\}
\end{equation}
where $\lambda$ is a positive regularization parameter. An estimate for $L$ can then be obtained by:
\begin{equation}\label{eq10}
L\approx\frac{1}{N}\sum_{i=1}^{N}\|{x_{i}}^{(0)}\|_{0}
\end{equation}
where ${x_{i}}^{(0)}$ represents the $i^{th}$ column of $X^{(0)}$.

We develop an iterative method to solve Eq. (\ref{eq8}). The idea is to find $X, \hat{X}$ with a fixed values of $D, W$ and then obtain $D, W$ with the updated values of $X,\hat{X}$. This process is repeated until $D,W$ converge. Since, we have already obtained an initial value for $D$ from Eq. (\ref{eq9}), we need to find an initial value for $W$. To find an initial value for $W$, we obtain the sparse codes $X$ and $\hat{X}$ by solving the following equations:
  \begin{equation*}
  \arg\min_{\|X\|_{0} \leq L}\|Y - DX\|_{F}^{2};\mbox{    }\arg\min_{\|\hat{X}\|_{0} \leq L}\|\hat{Y} - D\hat{X}\|_{F}^{2}
  \end{equation*}
The above can be combined to find $\bar{X}$ in Eq. (\ref{eq11}) using orthogonal matching pursuit (OMP) \cite{tropp2007signal}.
  \begin{equation}\label{eq11}
  \arg\min_{\|\bar{X}\|_{0} \leq L}\|\bar{Y} - D\bar{X}\|_{F}^{2}
  \end{equation}
    where, $\bar{Y} = [Y \mbox{ }\hat{Y}]$, $\bar{X} = [X \mbox{ } \hat{X}]$. Then, to obtain the initial value for $W$, we use the method proposed in \cite{jiang2013label} which is given by:
  \begin{equation}\label{eq12}
  W = \bar{H}\bar{X}^{t}(\bar{X}\bar{X}^{t} + \lambda_{1}I)^{-1}
  \end{equation}
  where $\bar{H} = [H \mbox{ }\tilde{H}]$. $\lambda_{1}$ is a positive regularizer parameter. Once the initial value of $W$ is obtained, we construct the following vectors:
  $$ Y_{new} = \begin{pmatrix}Y \\ \sqrt{\beta}H\end{pmatrix}\mbox{ },\hat{Y}_{new}=\begin{pmatrix}\hat{Y} \\ \sqrt{\beta}\tilde{H}\end{pmatrix}, D_{new} = \begin{pmatrix}D \\ \sqrt{\beta}W \end{pmatrix}$$
  As we have the initial values of $D,W$, we obtain the values of $X, \hat{X}$ by solving the following equation:
     \begin{equation}\label{eqNew}
     \arg\min_{\|\bar{X}\|_{0} \leq L}\|\bar{Y}_{new} - D_{new}\bar{X}\|_{F}^{2}
  \end{equation}
  where $\bar{Y}_{new} = [Y_{new} \mbox{ }\hat{Y}_{new}]$, $\bar{X} = [X \mbox{ } \hat{X}]$.

  With these values of $X$ and $\hat{X}$, we find $D_{new}$ by solving the problem in Eq. (\ref{eq13}) which automatically gives the values for $D, W$.
  \begin{equation}\label{eq13}
  \arg\min_{D_{new}}\bigg\{\frac{1}{N}\|Y_{new}-D_{new}X\|_{F}^{2} - \frac{\rho}{\hat{N}}\|\hat{Y}_{new}-D_{new}\hat{X}\|_{F}^{2}\bigg\}
  \end{equation}

  Using the definition of Frobenius norm, the above equation expands to:
  \begin{equation}\begin{aligned}\label{eq13a}
  \arg\min_{D_{new}}&\bigg\{\frac{1}{N}(Y_{new}-D_{new}X)(Y_{new}-D_{new}X)^{T}\\
   &- \frac{\rho}{\hat{N}}(\hat{Y}_{new}-D_{new}\hat{X})(\hat{Y}_{new}-D_{new}\hat{X})^{T}\bigg\}
   \end{aligned}
  \end{equation}

Applying the properties of trace and neglecting the constant terms in Eq. (\ref{eq13a}), solution to the problem in Eq. (\ref{eq13}) is equivalent to
  \begin{equation}\label{eq14}
  \arg\min_{D_{new}}\{-2\mbox{trace}(ED_{new}^{T}) + \mbox{trace}(D_{new}FD^{T}_{new}))\}
  \end{equation}
  where, $E = \frac{1}{N}Y_{new}X^{T} - \frac{\rho}{\hat{N}}\hat{Y}_{new}\hat{X}^{T}$; $F = \frac{1}{N}XX^{T} - \frac{\rho}{\hat{N}}\hat{X}\hat{X}^{T}$.
  The problem in Eq. (\ref{eq14}) is convex if $F$ is positive semidefinite. However, $F$ is not guaranteed to be positive semidefinite. To make $F$ a positive semidefinite matrix, $\rho$ should be chosen in a way such that the following condition is met:
  \begin{equation}\label{eq15}
  \frac{1}{N}\lambda_{min}(XX^{T}) - \frac{\rho}{\hat{N}}\lambda_{max}(\hat{X}\hat{X}^{T}) > 0
  \end{equation}
  where $\lambda_{min}(\bullet)$ and $\lambda_{max}(\bullet)$ represent the minimum and maximum eigenvalues of the corresponding matrices.
  Once an appropriate $\rho$ is chosen, Eq. (\ref{eq14}) can be solved using dictionary update step in \cite{mairal2010online}. After we obtain $D_{new}$, Eq. (\ref{eqNew}) is solved again to obtain new values for $X$ and $\hat{X}$ and we keep iterating between these two steps to obtain the final $D_{new}$.  The entire procedure is formally described in Algorithm \ref{alg 2}, which is used on a per-class basis to learn 3 class/segment specific dictionaries corresponding to brain, CSF and subdurals.

    \begin{algorithm}[t]
\caption{FLIS algorithm}\label{alg 2}
\begin{algorithmic}[1]
\State \textbf{Input:} $Y$, $\hat{Y}$, $H$, $\rho$, $\beta$, dictionary size $K$
\State \textbf{Output:} $D$, $W$
\Procedure{FLIS}{}
\State Find $L$ and an initial value for $D$ using Eq. (\ref{eq9}) and Eq. (\ref{eq10})
\State Find $X$ and $\hat{X}$ using Eq. (\ref{eq11})
\State Initialize $W$ using Eq. (\ref{eq12})
\State Update $ Y_{new} = \begin{pmatrix}Y \\ \sqrt{\beta}H\end{pmatrix}\mbox{ },\hat{Y}_{new}=\begin{pmatrix}\hat{Y} \\ \sqrt{\beta}\tilde{H}\end{pmatrix}, D_{new} = \begin{pmatrix}D \\ \sqrt{\beta}W \end{pmatrix}$
\State Update $X, \hat{X}$ using Eq. (\ref{eqNew})
\While{\textbf{not converged}}
\State Fix $X$, $\hat{X}$ and calculate $E = \frac{1}{N}Y_{new}X^{T} - \frac{\rho}{\hat{N}}\hat{Y}_{new}\hat{X}^{T}$; $F = \frac{1}{N}XX^{T} - \frac{\rho}{\hat{N}}\hat{X}\hat{X}^{T}$ \State Update $D_{new}$ by solving
$$\arg\min_{D_{new}}\{-2\mbox{trace}(ED_{new}^{T}) + \mbox{trace}(D_{new}FD^{T}_{new}))\}$$
\State Fix $D_{new}$, find $X$ and $\hat{X}$ using Eq. (\ref{eqNew})
\EndWhile
\EndProcedure
\State \textbf{RETURN:} $D_{new}$
\end{algorithmic}
\end{algorithm}

After we obtain class specific dictionaries and linear classifiers, we concatenate them to obtain $D = [D_{B} \mbox{ }D_{F} \mbox{ }D_{S}]$ and $W = [W_{B} \mbox{ }W_{F} \mbox{ }W_{S}]$. \\
\noindent\textbf{Assignment of a test pixel to a class (segment):} Once the dictionaries are learned, to classify a new pixel $z$, we extract a patch of size $w\times w$ around it to collect the intensity values and distance values from the boundary of the brain for the elements in the patch to form column vector $m(z)$. Then we find the sparse code $\alpha$ in Eq. (\ref{eq16}) using the learned dictionary $D$. Once $\alpha$ is obtained, we classify the pixel using Eq. (\ref{eq17}).
\begin{equation}\label{eq16}
\arg\min_{\alpha > 0}\|m(z) - D\alpha\|_{2}^{2} + \lambda\|\alpha\|_{1}
\end{equation}
\vspace{-.3cm}
\begin{equation}\label{eq17}
H_{z} = W\alpha,\mbox{ }label = \arg\max_{j}(H_{z}(j))
\end{equation}

\vspace{-.25cm}
\subsection{Training and Test Procedure Design for Hydrocephalic Image Segmentation: }
\label{sec:implementation}
\vspace{-.1cm}
\noindent \textbf{Training Set-Up:} In selecting training image patches for segmentation, it is infeasible to extract patches for all the pixels in each training image because that would require a lot of memory. Further, it is desired that patches used from training images should be in correspondence with the patches from test images. For example, training patches collected from the slices in the middle of the CT stack cannot be used for segmenting a slice that belongs to top or bottom. To address this problem, we divide the entire CT-stack of any patient into $P$ partitions such that images belonging to a given partition are anatomically similar. For each image in a partition (i.e a sub collection of CT image stack),  we must carefully extract patches to have enough representation from the 3 classes (segments) and likewise have enough diversity in the range of distances from the boundary of the brain.

\begin{figure}
 \begin{center}
  \includegraphics[scale=.21]{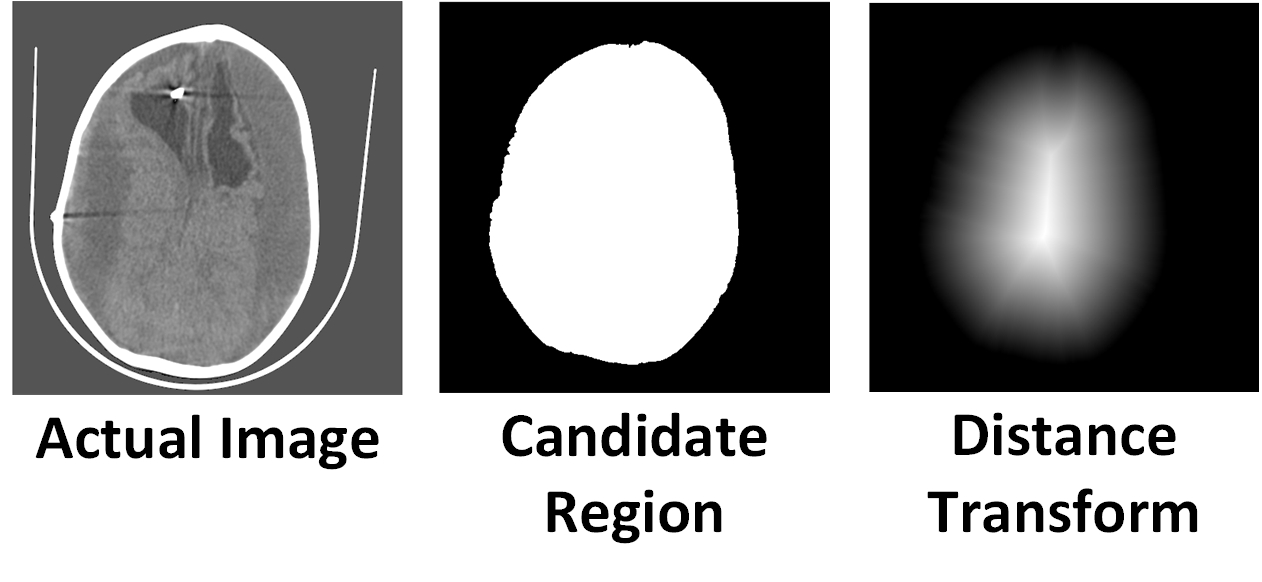}
 \end{center}
  \vspace{-.5cm}
  \caption{Visual representation of obtaining distance values from a CT-slice.}
  \label{Fig:distTrans}
  \vspace{-.4cm}
\end{figure}

\begin{figure}[t]
 \begin{center}
  \includegraphics[scale = .21]{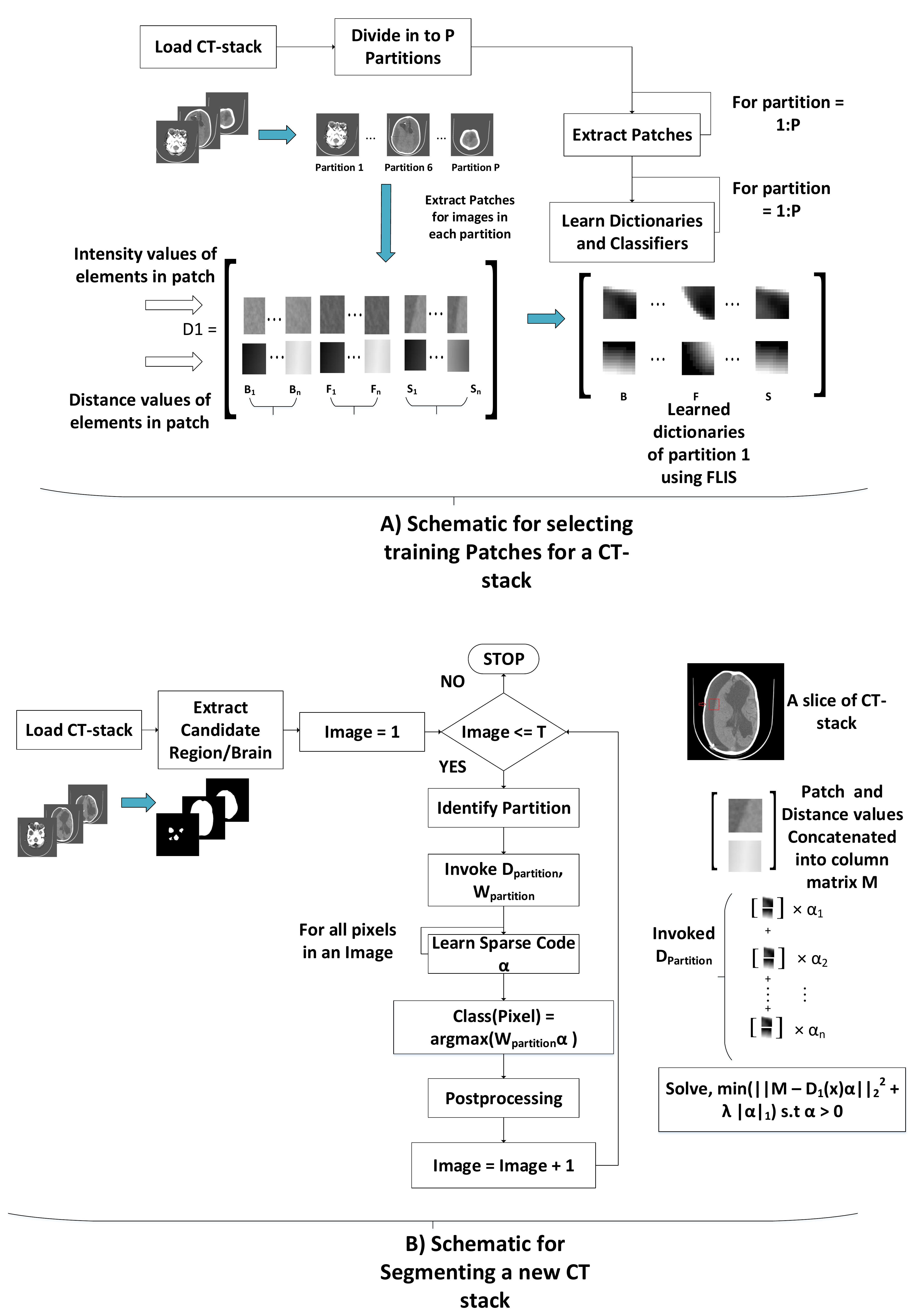}
  \end{center}
  \vspace{-.5cm}
  \caption{A) illustrates the procedure for selecting patches for training. B) illustrates the procedure for segmentation of a new CT- stack}\label{trainingSetup}
  \vspace{-.5cm}
\end{figure}

\noindent \textbf{Patch Selection Strategy for each class/segment:} First we find a {\em candidate} region for each image in the CT-stack by using an optical flow approach as mentioned in \cite{mandell2015volumetric}. The candidate region is a binary image which labels the region of an image that is to be segmented into brain, CSF and subdurals as 1. Then, the distance value for each pixel $z$ is given by $DT(z) = \min(d(z,q)) : CR(q)=0$, where $d(z,q)$ is the Euclidean distance between pixel $z$ and pixel $q$ and $CR$ is the candidate region. For a pixel $z$, it is essentially the minimum distance calculated from all the pixels that are not part of the candidate region. The candidate region of a sample image and its distance transform is shown in Fig. \ref{Fig:distTrans}. A subset of ``these distances" should be used in our training feature vectors. For this purpose, we propose a simple strategy wherein first we calculate the maximum and minimum distance of a given label/class in a CT image and pick patches randomly such that the distance range is uniformly sampled from min to max values.
The pseudo-code for this strategy and more implementation details can be found in \cite{PSU_venkat}.\\
Once training patches for each partition are extracted, we learn dictionaries and linear classifiers for each partition using the objective function described in Section \ref{sec:formulation}.The entire training setup and segmentation of a new test CT stack is summarized as a flow chart in Figure \ref{trainingSetup}.

\section{Experimental Results}
\label{sec:impRes}
\vspace{-.1cm}
We report results on a challenging real world data set of CT images acquired from the CURE Children\textquotesingle s Hospital of Uganda. Each patient (on an average) is represented by a stack of 28 CT images. We choose the number of partitions of such a stack $P$ to be 12 based on neurosurgeon feedback. The size of each slice is $512\times 512$. Slice thickness of the scans varied from 3mm to 10mm. The test set includes 15 patients while the number of training patients ranged from 9-17 and were non-overlapping with the test set. To validate our results, we used the dice-overlap coefficient, which for regions $A$ and $B$ is defined as

  \begin{equation}\label{eq18}
  DO(A,B) = \frac{2|A\cap B|}{|A| + |B|}
  \end{equation}
Note, $DO(A,B)$ evaluates to $1$, only when $ A = B$. The dice-overlap is computed for each method by using carefully obtained manually segmented results under the supervision of an expert neurosurgeon - (SJS). The proposed FLIS is compared against the following state of the art methods:
\begin{itemize}
\item SRC \cite{wright2009robust} based segmentation was implemented in \cite{wang2014integration} by using pre-defined dictionaries for each voxel/pixel in the scans. The objective function and classification procedure proposed in their work is implemented on our data set.
\item LC-KSVD \cite{jiang2013label} based dictionary learning method was used to segment MR brain images in \cite{tong2013segmentation} for hippocampus labeling. Two types of implementations were proposed in their paper which are named as DDLS and F-DDLS. In Fixed-DDLS (F-DDLS) dictionaries are learned offline and segmentation is performed online to improve speed of segmentation whereas in DDLS both operations are performed simultaneously. In this paper, we compare with the DDLS approach, as storing a dictionary for each pixel offline requires a very large memory.
\end{itemize}
Apart from these two methods, there are few others that use dictionary learning and a sparsity based framework for medical image segmentation \cite{nouranian2016learning, bevilacqua2016dictionary, lee2015brain, zhou2015nuclei, liao2013sparse, wu2014prostate,roy2015subject}. The objective function used in these aforementioned methods is similar to the above two methods with the application being different. We chose to compare against \cite{wang2014integration} and \cite{tong2013segmentation} because they are widely cited and were also applied to brain image segmentation.
\vspace{-.25cm}
\subsection{The need for a learning framework}
\label{sec:LearningAppr}
\vspace{-.1cm}
Before we compare our method against the state of the art in learning based segmentation, we demonstrate the superiority of the learning based approaches in comparison to the traditional intensity based methods. It was illustrated visually in Fig.\ \ref{TraditionalDemo} in Section \ref{sec:Intro} that intensity based methods find it  difficult to differentiate subdurals from brain and CSF. To validate this quantitatively, we compare dice-overlap coefficients obtained by using the segmentation results of \cite{bis2014}\footnote{Note that the method in \cite{bis2014} was implemented for MR brain images. We adapted their strategy for segmenting our CT images.} which is one of the best known intensity based methods and addressed as Brain Intensity Segmentation (BIS). The comparisons are reported in Table \ref{quantresults}. The learning based methods use a patch size of $11\times 11$ with number of training patients set to 15 and the sizes of individual class specific dictionaries set to 80.

\begin{figure}
 \begin{center}
  \includegraphics[scale=.23]{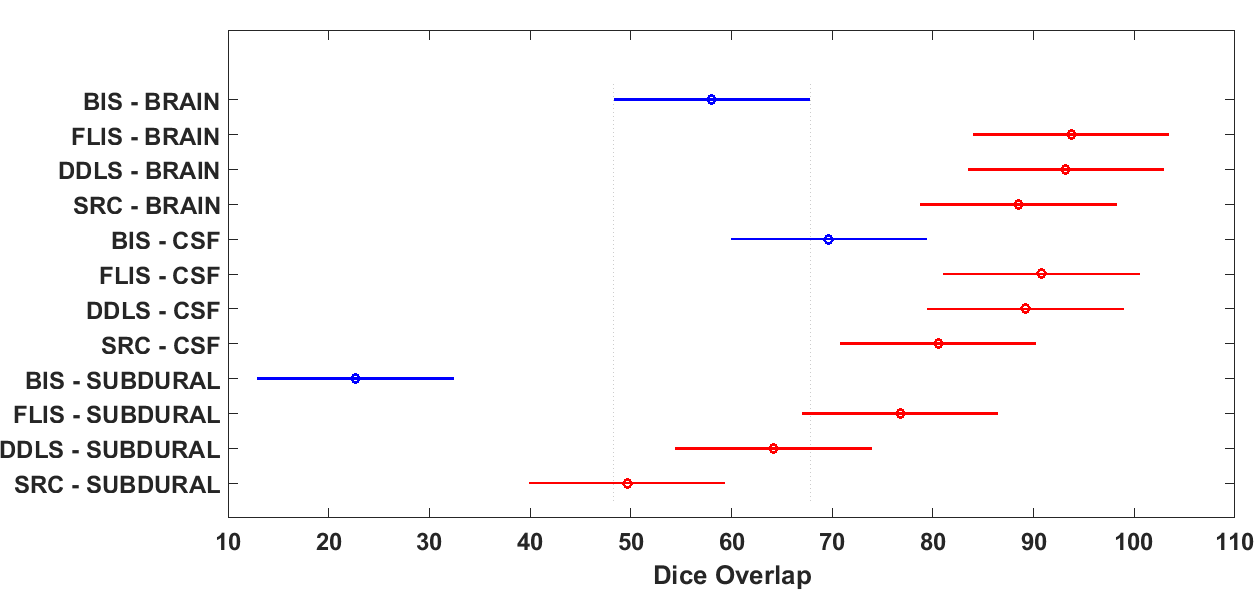}
 \end{center}
 \vspace{-.5cm}
  \caption{Comparison of traditional intensity based thresholding method with learning based approaches by a two-way ANOVA. Values reported by ANOVA across the method factor are $df = 3$, $F = 45.23$, $p \ll .01$, indicating that results of learning based approaches are significantly different and better than BIS. The intervals shown represent the 95 percent confidence intervals of the dice overlap values for the corresponding method-class configuration. Blue color represents BIS method and Red indicates the learning based approaches.}
   \label{Fig:AnovaTb1}
\end{figure}

The results in Table \ref{quantresults} confirm that learning based methods clearly outperform the traditional intensity based method, esp. in terms of the accuracy of identifying subdurals. Note that the dice overlap values in Table \ref{quantresults} for each class/segment are averaged over the 15 test patients. This will be the norm for the remainder of this Section unless otherwise stated. We performed a balanced two-way Analysis of Variance (ANOVA)\footnote{Prior to application of ANOVA, we rigourously verified that the observations (dice overlap values) satisfy ANOVA assumptions\cite{mcdonald2009handbook}.}  \cite{wu2011experiments} on the dice overlap values across patients for all $3$ classes (Brain, CSF and Subdural). Fig. \ref{Fig:AnovaTb1} illustrates these comparisons using posthoc Tukey range test \cite{wu2011experiments} and confirms that SRC, DDLS and FLIS (learning based methods) are significantly separated from BIS. $p$ values of BIS compared with other methods are observed to be much less than $.01$ which emphasizes the fact that learning based methods are more effective.

\begin{table}[t]
\caption{Comparison of learning based method with traditional intensity based thresholding method. Values are reported in Mean$\pm$SD(standard deviation) format}
\label{quantresults}
\vspace{-.5cm}
\begin{center}
\begin{tabular}{|c|c|c|c|}
\hline
\textbf{Method} & \textbf{Brain} & \textbf{CSF} & \textbf{Subdural} \\
\hline
\textbf BIS {\cite{bis2014}} & $.580\pm 0.21$ & $.696\pm 0.18$ & $.226 \pm 0.14 $\\
\hline
 \textbf{Patch based SRC \cite{wang2014integration}}& $.885 \pm 0.15$ & $.805 \pm 0.22$ & $.496 \pm 0.28$\\
\hline
\textbf{DDLS \cite{tong2013segmentation}} & $.932 \pm 0.04$ & $.892\pm 0.08$ & $.641\pm 0.2$\\
\hline
\textbf{FLIS (our method)} & $.937 \pm 0.02$ & $.908 \pm 0.07$ & $.767 \pm 0.14$ \\
\hline
\end{tabular}
\end{center}
\end{table}

\vspace{-.25cm}
\subsection{Parameter Selection:}
\label{sec:ParameterSelection}
\vspace{-.1cm}
In our method, several parameters have to be chosen carefully before we start implementation. Some of the important parameters are patch size, dictionary size, number of training patients and regularization parameters $\rho$ and $\beta$. $\rho$ and $\beta$ are picked by a cross-validation procedure \cite{kohavi1995study,lee2014efficient} such that $\rho$ is in compliance with Eq. (\ref{eq15}). The best values are found to be $\rho = .5$ and $\beta=2$. Our algorithm is fairly robust to other parameters such as patch size, number of training patients and length of dictionaries which is discussed in the subsequent sub-sections.
\vspace{-.25cm}
\subsection{Influence of Patch Size:}
\label{sec:PatchSize}
\begin{figure*}
 \begin{center}
  \includegraphics[scale=.20]{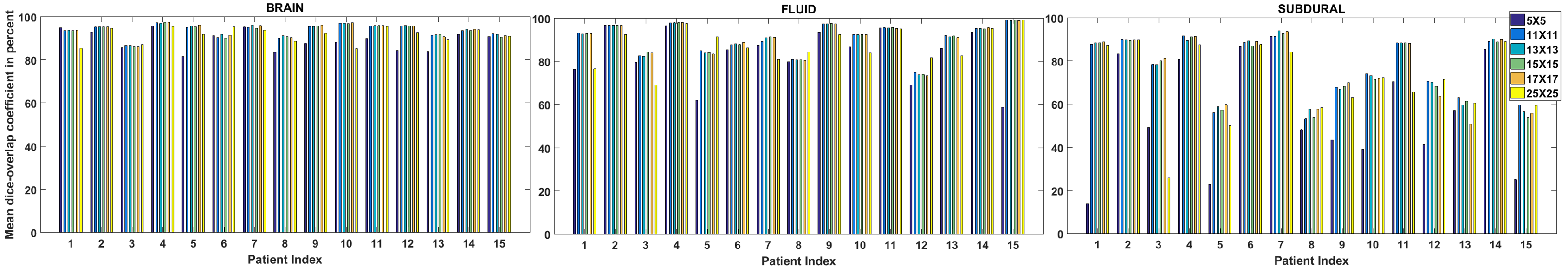}
 \end{center}
 \vspace{-.5cm}
  \caption{Mean dice overlap coefficients for all the 15 patients using our method are reported in this figure. Results for different square patch sizes varying from 5 to 25 are reported.}
   \label{Fig:resultsPatch}
\end{figure*}

\vspace{-.1cm}
If the patch size is very small, namely a single pixel in the extreme case, the necessary spatial information to accurately determine its class/segment is unavailable. On the other hand, a very large patch size might include pixels from different classes. For the experiment performed, the dictionary size of each class/segment and number of training patients for performing experiments are set to 120 and 17 respectively.
Experiments are reported for square patch windows with size varying from $5$ to $25$. The mean dice overlap values for all the 15 patients that are shown in Fig. \ref{Fig:resultsPatch} reveal that the results are quite stable for patch size in the range $11$ to $17$, indicating that while patch size should be chosen carefully, FLIS is robust against small departures from the optimal choice.

\vspace{-.25cm}
\subsection{Influence of Dictionary Size:}
\label{sec:DictSize}
\vspace{-.1cm}
Dictionary size is another important parameter in our method. Similar to patch size, very small dictionaries are incomplete and can not represent the data accurately. However, large dictionaries can represent the data more accurately, but at the cost of increased run-time and memory requirements.

In the results presented next, varying dictionary sizes of 20, 80, 120 and 150 are chosen. Note that these dictionary sizes are for each individual class. However, DDLS does not use class specific dictionaries. Therefore, to maintain consistency in both the methods, the overall dictionary size for DDLS is fixed to be 3 times the size of each individual dictionary in our method. Table \ref{DictSiz} compares FLIS with DDLS for different dictionary sizes. We did not compare with \cite{wang2014integration} as dictionary learning in not used in their approach. Experiments are conducted with a patch size of $13\times 13$ and with data from 17 patients used for training.

\begin{figure}
 \begin{center}
 \vspace{-.3cm}
  \includegraphics[scale=.25]{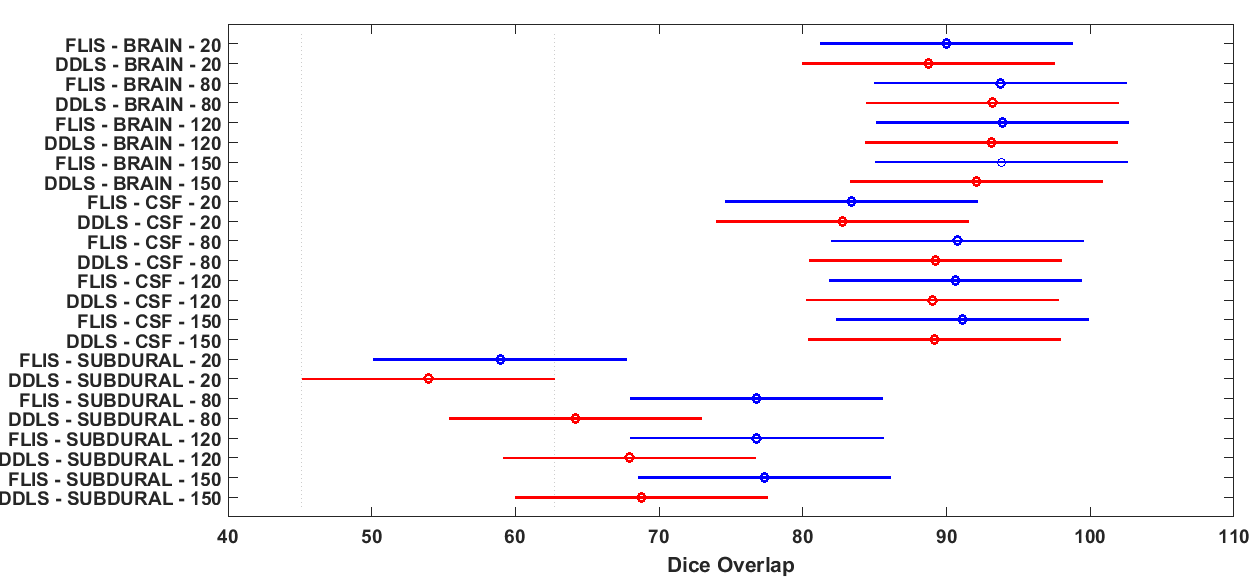}
 \end{center}
 \vspace{-.5cm}
  \caption{Comparison of FLIS with DDLS for different dictionary sizes by using a 3-way ANOVA. The intervals represent the 95 percent confidence intervals of dice overlap values for a given configuration of method-class-dictionary size. FLIS is represented in blue and DDLS in red. Values reported for ANOVA across the method factor are $df = 1$, $F = 7.22$, $p = .0075$. ANOVA values across dictionary length factor are $df = 3$, $F = 9.95$, $p\ll .01$. We also performed a repeated ANOVA across dictionary size factor for the two methods which reported a $p-$value=$1.73\times 10^{-10}$, which confirms that dictionary size has a significant role.}
   \label{Fig:dictStatRes}
\end{figure}

\begin{figure}
 \begin{center}
 \vspace{-.4cm}
  \includegraphics[scale=.18]{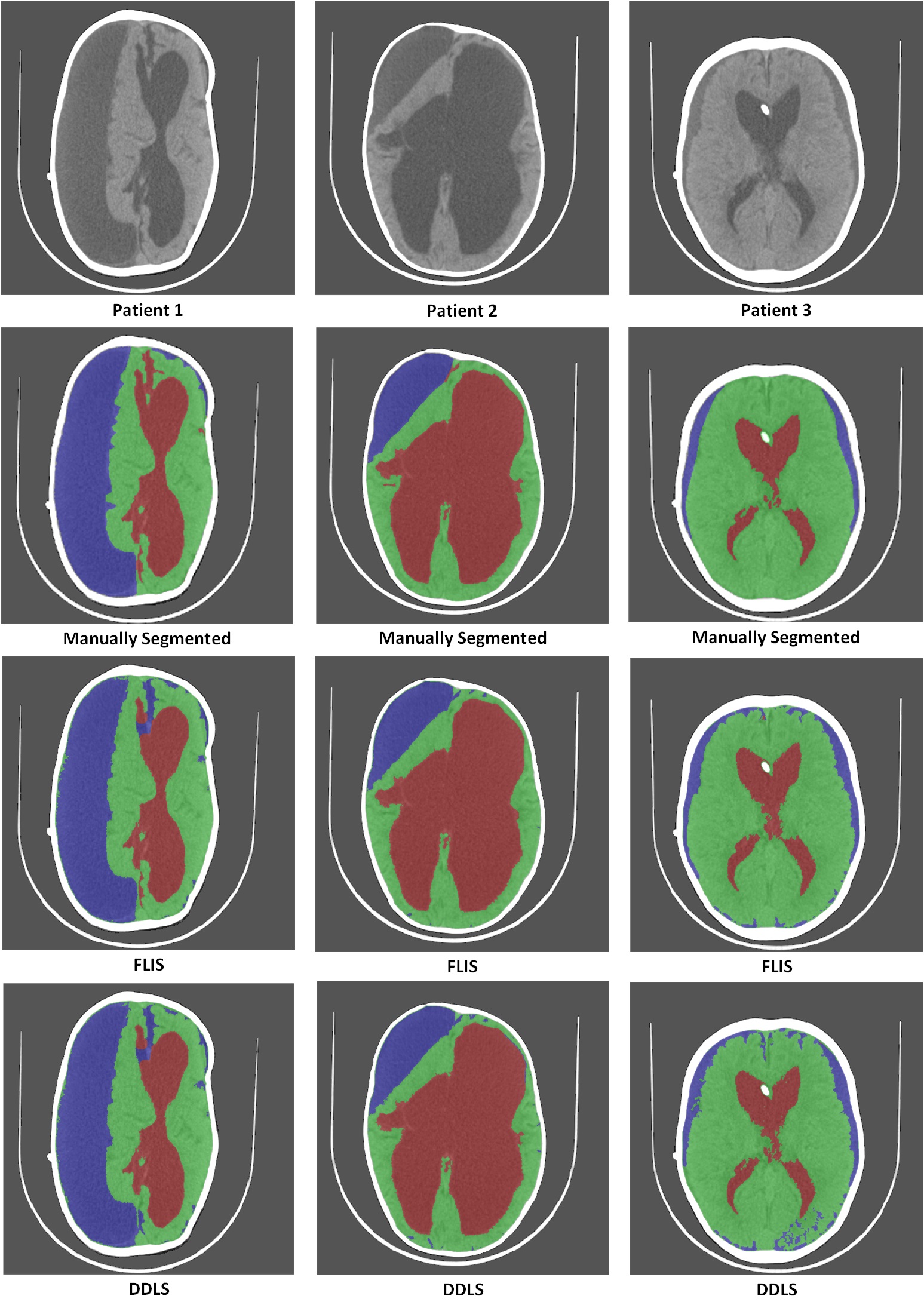}
 \end{center}
 \vspace{-.5cm}
  \caption{Comparison of results of the 2 methods for a dictionary size of 120 and training size of 17 patients. First row represents the original images of 3 patients. Second row represents their corresponding manually segmented image. Third row represents segmented images using FLIS. Fourth row represent segmented images using DDLS \cite{tong2013segmentation}. Green-Brain, Red-CSF, Blue-Subdurals.}
  \label{resultsDict}
\end{figure}

From Table \ref{DictSiz}, we observe that FLIS remains fairly stable with the change in size of dictionary whereas the DDLS method performed better in identifying subdurals as the size of dictionary is increased. For a fairly small dictionary size of $20$, the performance of both methods drops but FLIS is still relatively better. Further, to compare both the methods statistically, a 3-way balanced ANOVA is performed for all the 3 classes as shown in Fig. \ref{Fig:dictStatRes}. We observe that FLIS exhibits superior segmentation accuracy compared to DDLS although there is significant overlap between confidence intervals of FLIS and DDLS. This can be primarily attributed to the discriminative capability of the FLIS objective function which automatically discovers features that are crucial for separating segments. Visual comparisons are available in Figure \ref{resultsDict} when size of dictionary is set to 120. Visual results from Figure \ref{resultsDict} show that both the methods performed similarly in detecting large subdurals, but FLIS identifies subdurals more accurately in Patient 3 (3rd column of Fig.\ \ref{resultsDict}) where the subdurals have a smaller spatial footprint.

\begin{table}[t]
\caption{Performance of our method with different dictionary sizes. Values are reported in Mean$\pm$SD(standard deviation) format}
\label{DictSiz}
\vspace{-.5cm}
\begin{center}
\begin{tabular}{|c|c|c|c|c|}
\hline
\textbf{Dictionary size} & \textbf{Method} & \textbf{Brain} & \textbf{CSF} & \textbf{Subdural} \\
\hline
\multirow{2}{*}{\textbf{20}} & FLIS & $.891\pm 0.04$ & $.833 \pm 0.12 $ & $.580 \pm 0.23$ \\
                             & DDLS \cite{tong2013segmentation} & $.887 \pm 0.06$ & $.827 \pm 0.12$ & $.539 \pm 0.30$ \\
\hline
\multirow{2}{*}{\textbf{80}} & FLIS & $.939 \pm 0.03 $ & $.907 \pm 0.07 $ & $.770 \pm 0.13$ \\
                             & DDLS \cite{tong2013segmentation} & $.932\pm 0.05$ & $.892\pm 0.08$ & $.641\pm 0.26$ \\
\hline
\multirow{2}{*}{\textbf{120}} & FLIS & $.940 \pm 0.03$ & $.906 \pm 0.07 $ & $.768 \pm 0.14$ \\
                              & DDLS \cite{tong2013segmentation} & $.931\pm 0.04$ & $.890\pm 0.07$ & $.679\pm 0.17$ \\
\hline
\multirow{2}{*}{\textbf{150}} & FLIS &$.938 \pm 0.03 $ & $.911 \pm 0.07 $ & $.773 \pm 0.13 $\\
                              & DDLS \cite{tong2013segmentation} & $.921\pm 0.04 $ & $.891\pm 0.08 $ & $.687\pm 0.19 $ \\
\hline
\end{tabular}
\end{center}
\vspace{-.4cm}
\end{table}
\vspace{-.25cm}
\subsection{Performance variation against training}
\label{sec:TrainPat}
\begin{figure*}
 \begin{center}
   \includegraphics[scale=.2]{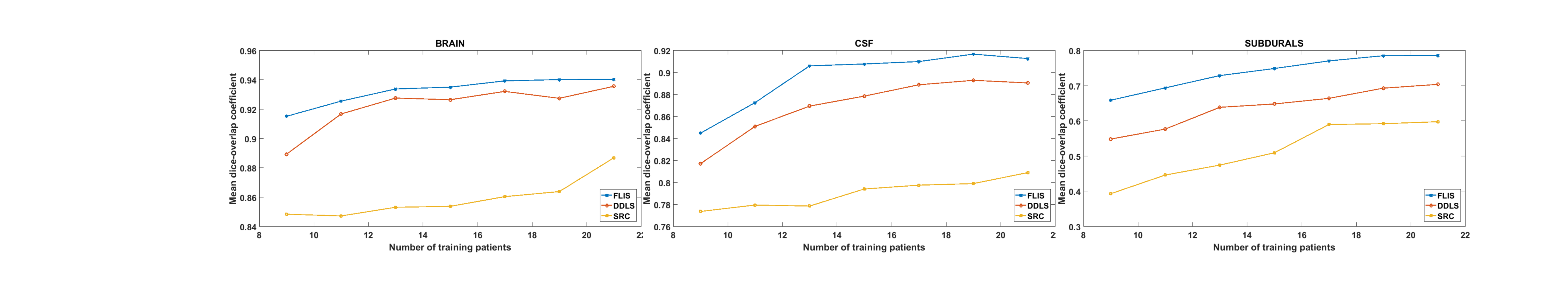}
 \end{center}
 \vspace{-.5cm}
  \caption{Comparing dice-overlap coefficients of FLIS with DDLS \cite{tong2013segmentation} and patch based SRC \cite{wang2014integration} for different sizes of training data. }
  \label{fig:trainData}
\end{figure*}

\vspace{-.1cm}
For the following experiment, we vary the number of training and test samples by dividing the total $32$ patients CT stacks into 9-23, 11-21, 13-19, 15-17, 17-15, 19-13 and 21-11 configurations (to be read as training-test). Figure \ref{fig:trainData} compares our method with DDLS and patch based SRC \cite{wang2014integration} for all these configurations. Note that, the results reported for each configuration are averaged over 10 random combinations of a given training-test configuration to remove selection bias.
The per-class dictionary size was fixed to $80$ for our method and DDLS, whereas for \cite{wang2014integration}, the dictionary size is determined automatically for a given training selection. The patch size is set to $13\times 13$.

A plot of dice overlap vs. training size is shown in Fig.\ \ref{fig:trainData}. Unsurprisingly, each of the three methods shows a drop in performance as the number of training image patches (proportional to the number of training patients) decreases. However, note that FLIS exhibits the most graceful degradation.

\begin{figure*}
 \begin{center}
  \includegraphics[scale=.21]{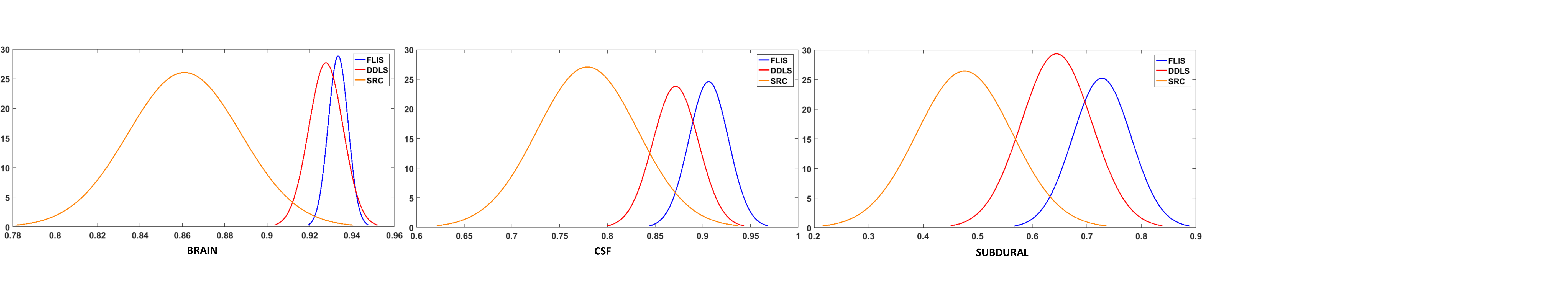}
 \end{center}
 \vspace{-.5cm}
  \caption{Gaussian fit for the histogram of dice overlap coefficients for ten random realizations of training data.}
  \label{fig:statTrain}
\end{figure*}

Fig.\ \ref{fig:statTrain} represents the gaussian fit for the histogram (for all 10 realizations combined) of dice-overlap coefficients for the configuration 13-19. Two trends may be observed: 1.) FLIS histogram has a mean higher than competing methods, indicating higher accuracy, 2.) the variance is smallest for FLIS confirming robustness to choice of training-test selection.

\begin{figure}[h]
 \begin{center}
 \vspace{-.4cm}
  \includegraphics[scale=.18]{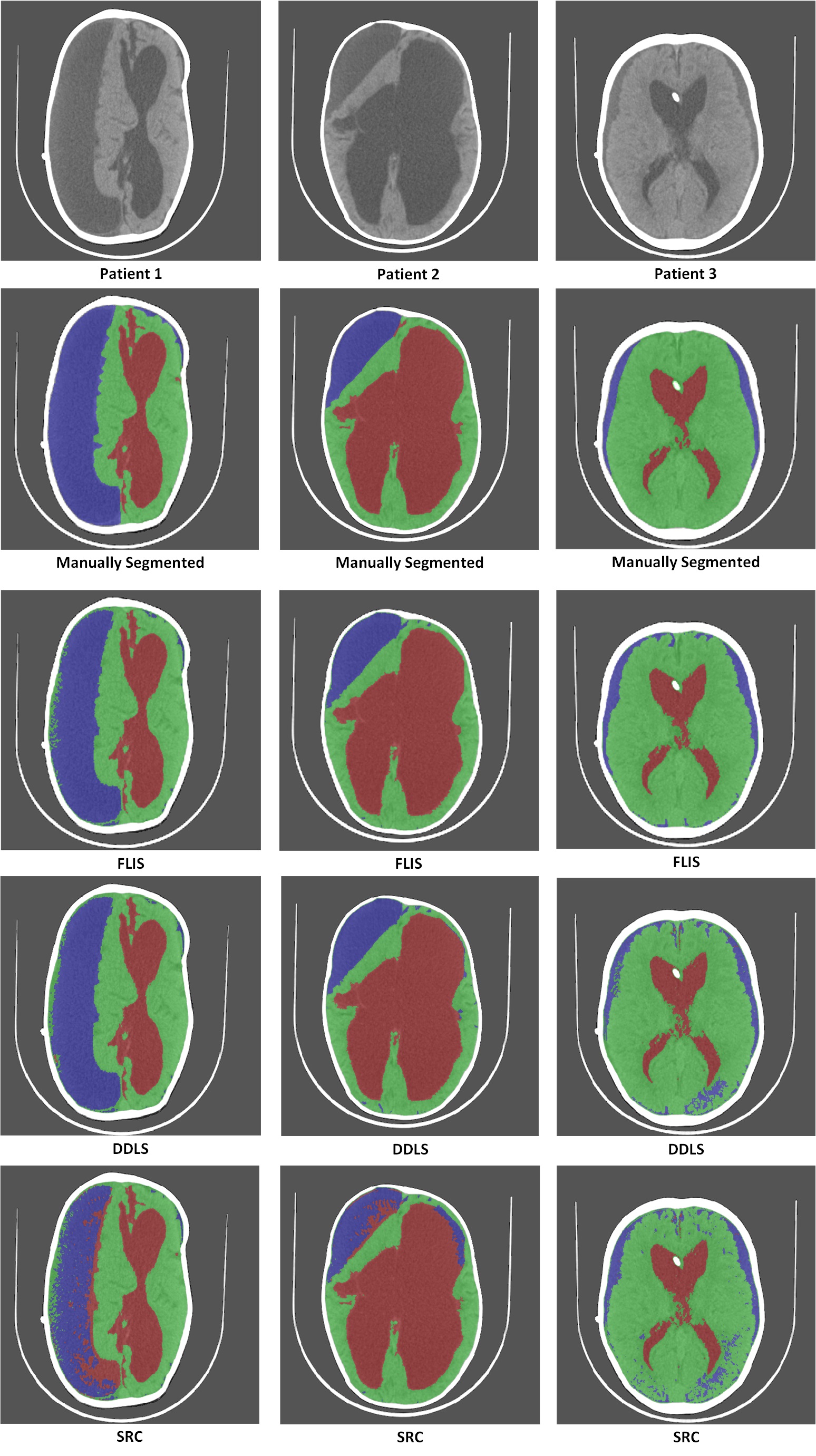}
 \end{center}
 \vspace{-.5cm}
  \caption{Comparison of results of the 3 methods for a training size of 17 patients. First row represents the original images of 3 patients. Second row represents their corresponding manually segmented image. Third row represents segmented images using FLIS. Fourth and Fifth rows represent segmented images using DDLS \cite{tong2013segmentation} and patch-based SRC \cite{wang2014integration} respectively. Green-Brain, Red-CSF, Blue-Subdurals.}
  \label{resultsTrain}
\end{figure}

Comparisons are visually shown in Figure \ref{resultsTrain}. A similar trend is also observed here where patch based SRC and DDLS improve as the number of training patients increase. We observe that DDLS and SRC based methods performed poorly in identifying the subdurals for Patient 3 (column 3) in Figure \ref{resultsTrain}. We also observe that both DDLS and FLIS outperform SRC implying that dictionary learning improves accuracy significantly.

\begin{figure*}
 \begin{center}
  \includegraphics[scale=.18]{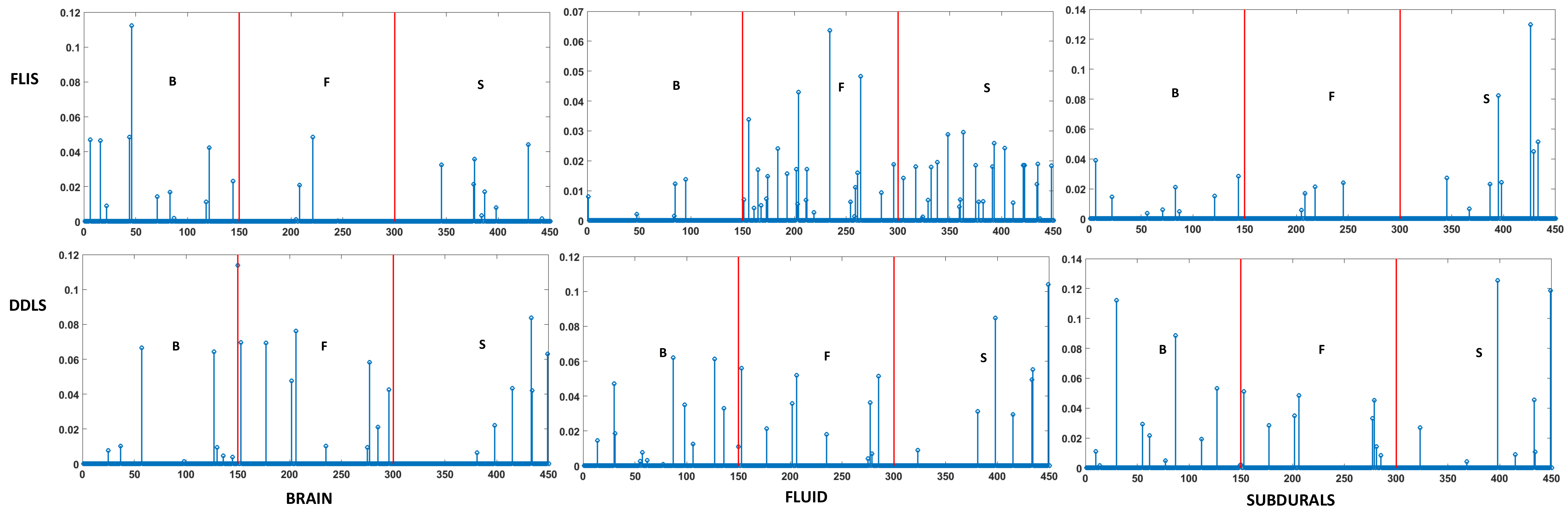}
 \end{center}
 \vspace{-.5cm}
  \caption{Comparing Sparse codes of a random pixel for brain (B), fluid (F) and subdurals (S). Row1: Sparse code for FLIS. Row2: Sparse code for DDLS. X axis indicates the dimension of the sparse codes. The left side of first red line correspond to brain, middle section corresponds to fluid and right side of second red line correspond to subdurals. Y axis indicate the values of the sparse codes.}
  \label{compSparse}
\end{figure*}
\vspace{-.25cm}
\subsection{Discriminative Capability of FLIS}
\label{sec:DiscriminativeCapability}
\vspace{-.1cm}
To illustrate the discriminative property of FLIS, we plot the sparse codes that are obtained from the classification stage for our method and DDLS for a single random pixel with a dictionary size of 150 in Fig.\ \ref{compSparse}. The two red lines in the figure act as a boundary for the 3 classes. For each of the three segments, i.e. brain, CSF and subdurals, we note that the active coefficients in the sparse code are concentrated more accurately in the correct class/segment for FLIS vs. DDLS.

To summarize the quantitative results, FLIS stands out particularly in its ability to correctly segment subdurals. The overall accuracy of brain and fluid segmentation is better than the accuracy of subdural segmentation for all the 3 methods. This is to be expected because the amount of subdurals present throughout in the images is relatively small compared to brain and fluid volumes.
\vspace{-.25cm}
\subsection{Computational Complexity}
\label{sec:CompComplex}
\vspace{-.1cm}
We compare the computational complexity of our FLIS with DDLS method. We do not compare with \cite{wang2014integration} as it does not learn dictionaries. Complexity of dictionary learning methods is estimated by calculating the approximate number of operations required for learning dictionaries for each pixel. Detailed derivation of complexity is presented in Appendix \ref{sec:Complexity}. The run-time and derived complexity per pixel are shown in Table \ref{estComp}. The run-time and computational complexity are derived per pixel. The values of parameters are defined as follows: The number of training patches $N = 4700$ for each class and the patch size is $11\times 11$. Sparsity level $L$ is chosen to be 5.  The run time numbers are consistent with the estimated number of operations shown in Table \ref{estComp} obtained by plugging in the values of above parameters in to the derived complexity formulas. FLIS is substantially less expensive from a computational standpoint. This is to be expected because DDLS uses pixel specific dictionaries, whereas FLIS dictionaries are class or segment specific but do {\em not} vary with the pixel location.

\vspace{-.25cm}
\subsection{Memory requirements}
\label{sec:Memory}

Memory requirements are derived in Appendix \ref{sec:MemoryReq}. The memory required for storing dictionaries for all the 3 methods are reported in Table \ref{MemoryAn}. These numbers are obtained assuming each element requires 16 bytes, and the following parameter choices: Number of training patients, $N_{t} = 15$, patch size = $11\times11$, $K = 80$ and $I_{x} = I_{y} = 512$. Consistent with Section \ref{sec:CompComplex}, the memory requirements of FLIS are also modest.

\begin{table}[t]
\caption{Complexity Analysis of methods}
\label{estComp}
\vspace{-.5cm}
\begin{center}
\begin{tabular}{|c|c|c|c|}
\hline
\textbf{Method} & \textbf{Complexity} & \textbf{Run time} & \textbf{Est. Operations}\\
\hline
\textbf{DDLS} & $\sim$ $9NK(2(\frac{d}{2} + 3) + L^{2})$ & $46.66$ seconds & $1.39\times 10^{9}$\\
\hline
\textbf{FLIS} & $\sim$  $\frac{9NK(2(d + 3) + L^{2})}{I_{x}\times I_{y}}$ & $.0003$ seconds & $1.005\times 10^{4}$  \\
\hline
\end{tabular}
\end{center}
\vspace{-.4cm}
\end{table}

\subsection{Comparison with deep learning architectures}
\label{sec:deep}
\begin{figure}
 \begin{center}
  \includegraphics[scale=.27]{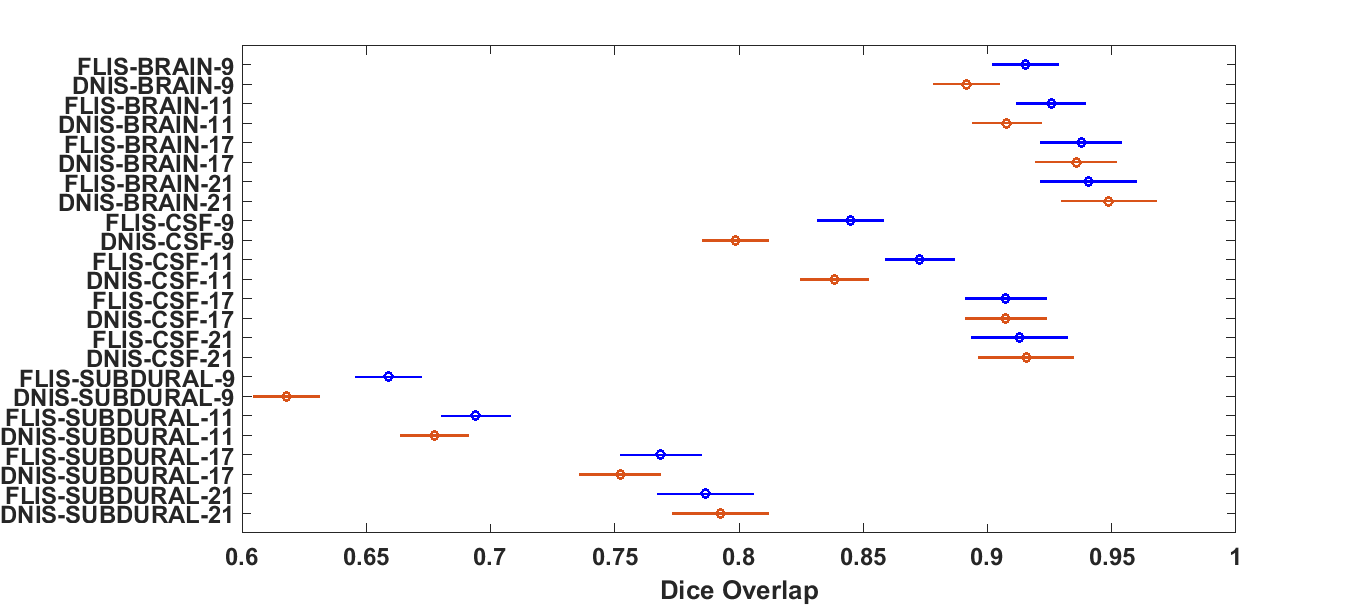}
 \end{center}
 \vspace{-.5cm}
  \caption{Comparison of FLIS with DNIS for different training configurations by using a 3-way ANOVA. The intervals represent the 95 percent confidence intervals of dice overlap values for a given configuration of method-class-training size. FLIS is represented in blue and DNIS in red. Values reported for ANOVA across the method factor are $df = 1$, $F = 35.54$, $p\ll .01$. ANOVA values across training size factor are $df = 3$, $F = 308.85$, $p\ll .01$.}
   \label{Fig:AnovaDeep}
\end{figure}

\begin{figure}
 \begin{center}
 \vspace{-.4cm}
  \includegraphics[scale=.18]{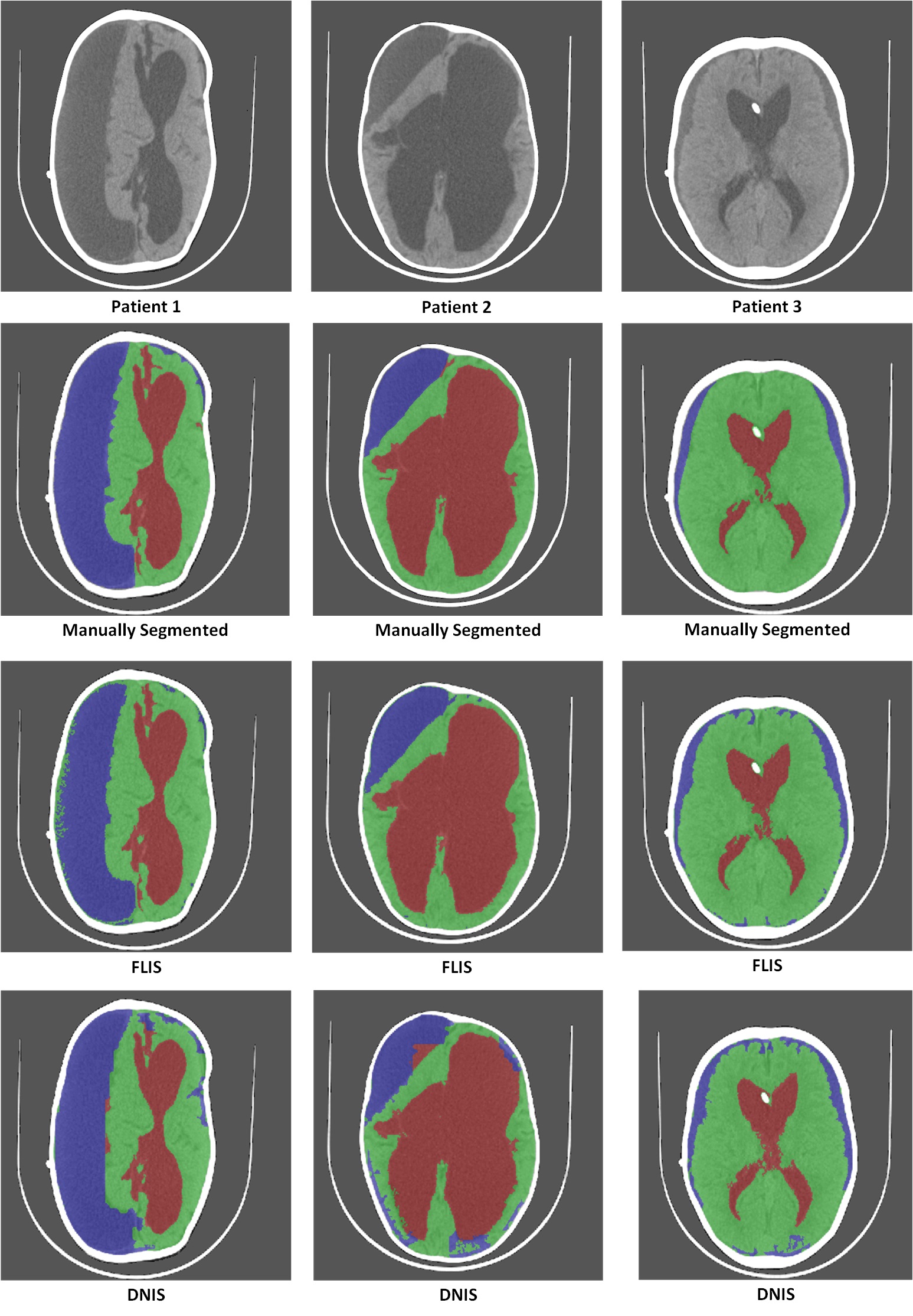}
 \end{center}
 \vspace{-.5cm}
  \caption{Comparison of results between DNIS and FLIS for training-test configuration of 17-15. First row represents the original images of 3 patients. Second row represents their corresponding manually segmented image. Third row represents segmented images using FLIS. Fourth row represent segmented images using DNIS. Green-Brain, Red-CSF, Blue-Subdurals.}
  \label{resultsDeepVis}
\end{figure}

A significant recent advance has been the development of deep learning methods, which have recently been applied to medical image segmentation \cite{moeskops2016automatic, pereira2016brain}. We implement the technique in \cite{moeskops2016automatic} which designs a convolutional neural network (CNN) for segmenting MR images. This method extracts 2D patches of different sizes centered around the pixel to be classified and a separate network is designed for each patch size. The output of each network is then connected to a single softmax layer to classify the pixel. Three different patch sizes were used in their work and the network configuration for each patch size is mentioned in Table \ref{tab:deepNet}. We reproduced the design in \cite{moeskops2016automatic} but with CT scans for training. We address this method as Deep Network for Image Segmentation (DNIS). Results in terms of comparisons with FLIS are shown in Table \ref{tab:deep}. Note that the training-test configuration of this experiment is the same as the one performed in subsection \ref{sec:TrainPat}. Unsurprisingly, FLIS performed better than DNIS for low training scenarios and DNIS performed slightly better than FLIS with an increase in number of training samples. Further, to confirm this statistically, a 3-way balanced ANOVA is performed for all the 3 classes as shown in Fig. \ref{Fig:AnovaDeep}. It may be inferred from Fig. \ref{Fig:AnovaDeep} that FLIS outperforms DNIS in the low to realistic training regime, while DNIS is competitive or mildly better than FLIS when training is generous. An example visual illustration of the results is shown for 3 patients in Fig. \ref{resultsDeepVis} where the benefits of FLIS are readily apparent. Also, note that the cost of training DNIS is in hours vs. the training time of FLIS which takes seconds -- see Table \ref{tab:deep} .

\begin{table}[t]
\caption{Memory requirements}
\label{MemoryAn}
\vspace{-.5cm}
\begin{center}
\begin{tabular}{|c|c|c|}
\hline
\textbf{Method} & \textbf{Memory(in bytes)} & \textbf{Approx Memory}\\
\hline
\textbf{SRC \cite{wang2014integration}}& $\frac{d}{2}\times \frac{d}{2}\times N_{t}\times I_{x}\times I_{y}\times 16$ & $\sim$ $9.2\times 10^{11}$ bytes \\
\hline
\textbf{DDLS \cite{tong2013segmentation}} & $(\frac{d}{2} + 3)\times 3K\times 16\times I_{x}\times I_{y}$ & $\sim$ $1.24\times 10^{11}$ bytes\\
\hline
\textbf{FLIS (our method)} & $(d + 3)\times 3K\times 16$ & $\sim$ $4.8\times 10^{5}$ bytes\\
\hline
\end{tabular}
\end{center}
\vspace{-.4cm}
\end{table}

\begin{table*}[tbh]
\caption{Deep network configuration of DNIS. Note: Conv- Convolutional layer followed by a $2\times 2$ Max pool layer, FC- Fully connected layer}
\label{tab:deepNet}
\vspace{-.5cm}
\begin{center}
\begin{tabular}{|c|c|c|c|c|}
\hline
\textbf{Patch Size} & \textbf{Layer1 (Conv)} & \textbf{Layer2 (Conv)} & \textbf{Layer3 (Conv)} & \textbf{Layer4 (FC)}\\
\hline
\textbf{$25\times 25$} & 24 $5\times 5\times 1$ & 32 $3\times 3\times 24$ & 48 $3\times 3\times 32$ & 256 nodes\\
\hline
\textbf{$50\times 50$} & 24 $7\times 7\times 1$ & 32 $5\times 5\times 24$ & 48 $3\times 3\times 32$ & 256 nodes\\
\hline
\textbf{$75\times 75$} & 24 $9\times 9\times 1$ & 32 $7\times 7\times 24$ & 48 $5\times 5\times 32$ & 256 nodes\\
\hline
\end{tabular}
\end{center}
\vspace{-.4cm}
\end{table*}

\begin{table*}[tbh]
\caption{Performance of our method with DNIS. Values are reported in Mean$\pm$SD(standard deviation) format}
\label{tab:deep}
\vspace{-.5cm}
\begin{center}
\begin{tabular}{|c|c|c|c|c|c|}
\hline
\textbf{Training samples} & \textbf{Method} & \textbf{Brain} & \textbf{CSF} & \textbf{Subdural} & \textbf{Training Time (in seconds)}\\
\hline
\multirow{2}{*}{\textbf{9}} & FLIS & $.915\pm 0.03$ & $.845 \pm 0.08 $ & $.660 \pm 0.14$ & $69.83$\\
                             & DNIS  & $.890 \pm 0.03$ & $.80 \pm 0.09$ & $.632 \pm 0.13$ & $2860.66$\\
\hline
\multirow{2}{*}{\textbf{11}} & FLIS & $.926 \pm 0.02 $ & $.873 \pm 0.07 $ & $.694 \pm 0.13$ &$96.61$ \\
                             & DNIS & $.910\pm 0.03$ & $.834\pm 0.07$ & $.671\pm 0.13$ & $9464.34$\\
\hline
\multirow{2}{*}{\textbf{13}} & FLIS & $.934 \pm 0.02$ & $.906 \pm 0.06 $ & $.729 \pm 0.14$ & $106.15$\\
                              & DNIS & $.919\pm 0.02$ & $.880\pm 0.07$ & $.690\pm 0.13$ & $10443.57$\\
\hline
\multirow{2}{*}{\textbf{15}} & FLIS &$.935 \pm 0.02 $ & $.908 \pm 0.06 $ & $.750 \pm 0.12 $ & $115.23$\\
                              & DNIS & $.934\pm 0.02 $ & $.897\pm 0.06 $ & $.728\pm 0.12 $ & $11823.99$\\
\hline
\multirow{2}{*}{\textbf{17}} & FLIS &$.939 \pm 0.02 $ & $.910 \pm 0.06 $ & $.770 \pm 0.11 $ & $124.41$\\
                              & DNIS & $.939\pm 0.02 $ & $.908\pm 0.05 $ & $.752\pm 0.12 $ & $12940.41$\\
\hline
\multirow{2}{*}{\textbf{19}} & FLIS &$.940 \pm 0.02 $ & $.917 \pm 0.06 $ & $.786 \pm 0.13 $ & $138.71$\\
                              & DNIS & $.943\pm 0.02 $ & $.914\pm 0.04 $ & $.786\pm 0.10 $ & $14669.76$ \\
\hline
\multirow{2}{*}{\textbf{21}} & FLIS &$.940 \pm 0.01 $ & $.913 \pm 0.04 $ & $.786 \pm 0.10 $ & $149.05$\\
                              & DNIS & $.950\pm 0.02 $ & $.919\pm 0.04 $ & $.792\pm 0.10 $ & $15846.87$ \\
\hline
\end{tabular}
\end{center}
\vspace{-.4cm}
\end{table*}

\section{Discussion and Conclusion}
\label{sec:Conc}

In this paper, we address the problem of segmentation of post-op CT brain images of hydrocephalic patients from the viewpoint of dictionary learning and discriminative feature discovery. This is very challenging problem from the distorted anatomy and subdural hematoma collections on these scans. This makes subdurals hard to differentiate from brain and CSF. Our solution involves a sparsity constrained learning framework wherein a dictionary (matrix of basis vectors) is learned from pre-labeled training images. The learned dictionaries under a new criterion are shown capable of  yielding superior results to state of the art methods. A key aspect of our method is that only class or segment specific dictionaries are necessary (as opposed to pixel specific dictionaries), substantially reducing the memory and computational requirements.

Our method was tested on real patient images collected from CURE Children\textquotesingle s Hospital of Uganda and the results outperformed well-known methods in sparsity based segmentation.


\appendices
\section{Complexity analysis}
\label{sec:Complexity}
We derive the computational complexity of our FLIS and compare it with DDLS\cite{tong2013segmentation}. Computational complexity for each method is derived by finding the approximate number of operations required  per pixel in learning the dictionaries. To simplify the derivation, let us assume that number of training samples and size of dictionary be same for all the 3 classes. Let they be represented as $N$ and $K$. Let us also assume that sparsity constraint $L$ remains the same for all the classes. Let the training samples be represented as $Y$ and the sparse code be represented as $X$.\\
Two major steps in most of the dictionary learning methods are the dictionary update and sparse coding steps, which in our case are $l_{0}$ minimization. The dictionary update step is solved either by using block coordinate descent \cite{mairal2010online} or the singular value decomposition \cite{aharon2006rm}. The second step which involves solving an Orthogonal Matching Pursuit \cite{tropp2007signal} is the most expensive step. Therefore, to derive the computational complexities, we find the approximate number of operations required to solve the sparse coding step in each iteration.
\vspace{-.25cm}
\subsection{Complexity of FLIS:}
\vspace{-.1cm}
As discussed above, we find the approximate number of operations required to solve the sparse coding step in our algorithm. To do that, first we find the complexity of the major sparse coding step which is given by Eq. (\ref{eq19}).

  \begin{equation}\label{eq19}
  \arg\min_{\|X\|_{0} \leq L}\|Y - DX\|_{F}^{2}
  \end{equation}
  where the dimension of $Y$ is equal to $\mathbb{R}^{d\times N}$ and dimension of $D$ is equal to $\mathbb{R}^{d\times K}$. For a batch-OMP problem with the above dimensions, the computational complexity is derived in \cite{rubinstein2008efficient} and it is equal to $N(2dK + L^{2}K + 3LK + L^{3}) + dK^{2}$. Assuming $L\ll K\approx d \ll N$, it approximately simplifies to
     \begin{equation}\label{eq20}
NK(2d + L^{2}).
  \end{equation}
The sparse coding step in our FLIS algorithm requires us to solve  $\arg\min_{\|\bar{X}\|_{0} \leq L}\|\bar{Y}_{new} - D_{new}\bar{X}\|_{F}^{2}$ where $\bar{Y}_{new} \in \mathbb{R}^{(d + 3)\times 3N}$ and $D_{new} \in \mathbb{R}^{(d + 3)\times K}$ which can be solved from Eq. (\ref{eqNew}). Substituting these values into Eq. (\ref{eq20}), we get the complexity of learning dictionary for a single class as $3NK(2(d + 3) + L^{2})$. Since we have 3 classes, the overall complexity of learning is multiplied by 3: $C_{FLIS} = 9NK(2(d + 3) + L^{2})$. As the same dictionary is used for all the pixels in an image $I$ with dimension $I_{x}\times I_{y}$ , $C_{FLIS} = \frac{9NK(2(d + 3) + L^{2})}{I_{x}\times I_{y}}$.
\vspace{-.25cm}
\subsection{Complexity of DDLS \cite{tong2013segmentation}:}
\vspace{-.1cm}
We already showed that by removing the discriminating term from FLIS in Eq. (\ref{eq7}), it turns into the objective function described for DDLS in Section \ref{sec:formulation}. Therefore, the most complex step remains the same for DDLS as well. However, since DDLS does not include distance feature the size of $d$ changes to $\frac{d}{2}$ and also it computes the dictionaries for all the classes at once. Keeping these two differences in mind, the computational complexity of DDLS is: $C_{DDLS} = 9NK(2(\frac{d}{2} + 3) + L^{2})$. In addition, a separate dictionary is computed for each pixel in DDLS, which means the complexity scales with the size of the image.

\section{Memory Requirements: }
\label{sec:MemoryReq}
We now calculate the memory required for our method and compare it with DDLS \cite{tong2013segmentation} and patch based SRC \cite{wang2014integration}. Memory requirement for all the methods is calculated by estimating the number of bytes required to store the dictionaries. In the case of FLIS and DDLS, the size of the dictionary plays an important role in calculating memory requirement whereas in SRC, the number of training images plays an important role as it uses pre-defined dictionaries. Another point to note is, as the entire CT stack is divided into $P$ partitions and a dictionary is stored for each partition, we derive the memory required for storing dictionaries for each individual partition. To obtain the total memory required, the formulas derived in the subsequent sections have to be multiplied by $P$.
\vspace{-.25cm}
\subsection{Memory required for FLIS: }Suppose the length of each dictionary is $K$ and the size of the column vector is $d$, then the size of the complete dictionary for all the 3 classes combined is $d\times 3K$. Further, we also store linear classifier $W$ for classification which is of size $3\times 3K$. Therefore, the complete size of the dictionary is $(d + 3)\times 3K$. Assuming each element in dictionary is represented by 16 bytes, the total memory in bytes required for storing FLIS dictionaries is $M_{FLIS} = (d + 3)\times 3K\times 16$.
\vspace{-.25cm}
\subsection{Memory required for DDLS \cite{tong2013segmentation}: }One major difference between FLIS and DDLS is the size of the column vector in DDLS is approximately half of the size in FLIS's case as the distance values are not considered in DDLS. The other major difference is a dictionary is stored for each individual pixel. Keeping these two differences in mind and with the same dictionary length, the total memory in bytes required for storing DDLS dictionaries is $M_{DDLS} = (\frac{d}{2} + 3)\times 3K\times 16\times I_{x}\times I_{y}$ where $I_{x}\times I_{y}$ is the image size.
\vspace{-.25cm}
\subsection{Memory required for Patch based SRC \cite{wang2014integration}: }In SRC method, predefined dictionaries for each pixel are stored instead of compact dictionaries. For a given pixel $x$ in an image, a patch of size $w\times w$ is considered around the same pixel location in training images and then a patch of size $w\times w$ around new pixels form the dictionary of pixel $x$. Assuming there are $N_{t}$ training images, the total size of the dictionary for a given pixel is $\frac{d}{2}\times \frac{d}{2}\times N$ as the size of the patch in this method is approximately half of the size of column vector in FLIS method. Therefore, the total memory in bytes required for this methods is $M_{SRC} = \frac{d}{2}\times \frac{d}{2}\times N_{t}\times I_{x}\times I_{y}\times 16$.
\section*{Acknowledgment}

We thank Tiep Huu Vu for providing his valuable inputs to this work supported by NIH grant R01HD085853.
%
%

%
%
%
%
%
%

\bibliographystyle{IEEEtran}
\bibliography{Refs}

\end{document}